\documentclass[aps,twocolumn,prc,superscriptaddress,noshowpacs,nofootinbib,noshowkeys,floatfix]{revtex4}

\usepackage[dvips]{graphics,graphicx}
\usepackage[colorlinks=true,linktocpage=true,linkcolor=blue,citecolor=blue]{hyperref}
\usepackage[usenames,dvipsnames]{color}
\usepackage{amsmath, amssymb}
\usepackage{multirow}
\usepackage{longtable}
\usepackage{color}
\usepackage{cleveref}
\usepackage[normalem]{ulem}  

\renewcommand\sout{\bgroup \color{blue} \ULdepth=-.5ex \ULset}

\begin{document}

\title{{\Large Dependence on beam energy and nuclear equation of state of anisotropic flow and particle production in low-energy heavy-ion collisions}}

\author{Sumit Kumar Kundu}
\affiliation{Discipline of Physics, School of Basic Sciences, Indian Institute of Technology Indore, Indore 453552 India}
\author{Yoshini Bailung}
\affiliation{Discipline of Physics, School of Basic Sciences, Indian Institute of Technology Indore, Indore 453552 India}
\author{Sudhir Pandurang Rode}
\affiliation{Discipline of Physics, School of Basic Sciences, Indian Institute of Technology Indore, Indore 453552 India}
\email{sudhirrode11@gmail.com}
\author{Partha Pratim Bhaduri}
\affiliation{Variable Energy Cyclotron Centre, HBNI, 1/AF Bidhan Nagar, Kolkata 700 064, India}
\author{Ankhi Roy}
\affiliation{Discipline of Physics, School of Basic Sciences, Indian Institute of Technology Indore, Indore 453552 India}
\date{\today}

\begin{abstract}
We analyse various flow coefficients of anisotropic momentum distribution of final state particles in mid-central ($b$ $=$ 5--9 $fm$) Au +  Au collisions in the beam energy range $\rm E_{\rm Lab}$ $=$ $1A -158A$ GeV. Different variants of the Ultra-relativistic Quantum Molecular Dynamics (UrQMD) model, namely the pure transport (cascade) mode and the hybrid mode, are employed for this investigation. In the hybrid UrQMD model, the ideal hydrodynamical evolution is integrated with the pure transport calculation for description of the evolution of the fireball. We opt for the different available equations of state (EoS) replicating the hadronic as well as partonic degrees of freedom together with possible phase transitions, viz. hadron gas, chiral + deconfinement EoS and Bag Model EoS, to investigate their effect on the properties of the final state particles. We also attempt to gain insights about the dynamics of the medium by studying different features of particle production such as particle ratios and net-proton rapidity distribution. The results and conclusions drawn here would be useful to understand the response of various observables to the underlying physics of the model as well as to make comparisons with the upcoming measurements of the future experiments at Facility for Antiproton and Ion Research (FAIR) and Nuclotron-based Ion Collider fAcility (NICA). 

\end{abstract}

\maketitle

\section*{I Introduction}
One of the main objectives of the modern day relativistic heavy-ion physics research is to understand the phase structure of the strongly interacting matter at extreme conditions of temperatures and net baryon densities in the laboratory~\cite{a,c}. Possible existence of critical point of QCD matter along with phase transition to deconfined state motivates the high energy community to continue the efforts in this direction. Exploration of the QCD matter at finite baryon densities is relatively less extensive compared to the one created at negligible baryon densities. Ample amount of investigations have been performed in the latter direction in past two decades with various experiments at Relativistic Heavy Ion Collider (RHIC)~\cite{rhic1,rhic2} and Large Hadron Collider (LHC)~\cite{lhc1,lhc2,lhc3}. Upcoming experiments at future accelerator facilities such as Nuclotron-based Ion Collider fAcility (NICA)~\cite{Kekelidze:2016wkp} and Facility for Antiproton and Ion Research (FAIR)~\cite{Ablyazimov:2017guv,Sturm:2010yit} aim to probe the baryon rich matter with good precision. However an optimal use of these facilities demand an extensive analysis of the available data and model based studies of different observables in the similar energy domain.

The anisotropic flow of the particles emitted in non-central relativistic heavy-ion collisions is considered as a promising observable to investigate the collective effects of the produced medium. Originated due to the pressure gradient as a result of the multiple scatterings among the constituents of the medium, it is vulnerable to the underlying nuclear equation of state. Azimuthal anisotropy in momentum distribution of the final state particles is quantified in terms of various harmonic coefficients using the Fourier series. These different anisotropic flow coefficients can be expressed as,
\begin{align*}
v_{n} = <\cos[n(\phi - \Psi)]>
\end{align*} 
where azimuthal angle of the particle and reaction plane angle are indicated by $\phi$ and $\Psi$, respectively. Moreover, $v_{n}$ is defined as directed flow ($v_{1}$), elliptic flow ($v_{2}$), triangular flow ($v_{3}$), quadrangular flow ($v_{4}$) for n $=$ 1, 2, 3, 4 and so on, respectively. These coefficients are believed to provide an insight on dynamics of the fireball. For instance, significant magnitude of the $v_{2}$ has shed light on the possibility that the bulk of the produced matter achieve close to local thermal equilibrium conditions. The pressure gradient developed due to rescatterings in the early stage of the collisions converts the initial state spatial anisotropy to final state momentum anisotropy and $v_2$. Several experiments~\cite{Alver:2006wh,Aamodt:2010pa} at different energies have examined $v_2$ for the possible signature of thermalization of the produced medium. Substantial amount of study has been performed to inspect $v_{2}$ in low energy collisions at various beam energy ranges~\cite{Bhaduri:2010wi,Sarkar:2017fuy,Auvinen:2013sba} availing variety of microscopic transport models~\cite{Bass:1998ca,Bleicher:1999xi,Lin:2004en,Chen:2004vha}. At low beam energies, change of sign, i.e. transition from out-of-plane to in-plane flow has been observed~\cite{LeFevre:2016vpp,Pinkenburg:1999ya}.

On the other hand, the directed flow, $v_1$, quantifies the deflection of the produced particles in the reaction plane. Sensitivity to the longitudinal dynamics and possibility of being developed prior to $v_2$~\cite{Nara:2016phs,Nara:1999dz,Konchakovski:2014gda}, make $v_1$ worth studying in relativistic nuclear collisions. The magnitude of $v_1$ is expected to vanish in the vicinity of the phase transition due to softening of the underlying EoS and this makes it an exciting observable for the analysis at RHIC-BES, FAIR and NICA energies. Plethora of the activities has been carried out in this direction in a past few decades at various experiments. For instance, the slope of $v_1$ being the measure of the signal strength, shows linearity at the midrapidity at AGS~\cite{Liu:2000am,Chung:2000ny,Chung:2001je} energies and below. However, this linearity at midrapidity is not expected to be maintained at higher beam energies because of the slope at midrapidity is found to be different than that at beam rapidity at energies above SPS~\cite{Appelshauser:1997dg,Adams:2004bi,Back:2005pc}. Hydrodynamical model calculations indicate that the so called structure "wiggle" is sensitive to the underlying EoS~\cite{Snellings:1999bt,Csernai:1999nf,Brachmann:1999xt}. Study of higher order harmonics has gained some attention in a past few years and expected to provide further insights about the produced fireball. Fourth order harmonic coefficient, $v_{4}$ has been known to be sensitive to intrinsic $v_{2}$~\cite{Borghini:2005kd,Gombeaud:2009ye,Luzum:2010ae} and therefore, it is quite interesting to investigate it over wider range of the beam energies which has also been attempted using microscopic transport model, JAM~\cite{Nara:1999dz,Nara:2018ijw}. It bears some crucial details about the collision dynamics predicted by hydrodynamical calculations~\cite{Luzum:2010ae}. 

In this article, we make some efforts to address the nuclear equations of state dependence of the anisotropic flow coefficients and particle production in non-central ($b$ = 5--9 $fm$) Au--Au collisions in very wide ranges of the beam energies, $\rm E_{\rm Lab}$ $=$ $1A -158A$ GeV which span over existing GSI-SIS energy of HADES experiment up to top SPS energy. It is found that the corresponding $<N_{\rm part}>$ values in the chosen impact parameter range $b$ = 5--9 $fm$, covers approximately $10-40\%$ centrality class~\cite{Adare:2015bua}. For our study, we employ the publicly available version 3.4 of the UrQMD model with different configurations of hybrid model for the intermediate hydrodynamical stage viz., Hadron Gas (HG), Chiral + deconfinement EoS and Bag Model EoS along with pure transport approach. The latter two hybrid versions mimic the partonic degrees of freedom and phase transition in the medium, however, the first one includes hadronic degrees of freedom only.  The reaction plane angle ($\Psi$) is taken to be zero within this model.  It is important to note that the present study is not a pioneering attempt to apply hybrid UrQMD model to study collective flow excitations at these beam energies. In~\cite{Petersen:2006vm}, the authors have calculated the transverse momentum and rapidity dependence of $v_1$ and $v_2$ at 40A and 160A GeV in Pb + Pb collisions using standard UrQMD model at various centralities which showed disagreement with experimental measurements by NA49 collaboration. In addition, $v_1$ and $v_2$ were also studied as a function of beam energy in the range of $\rm E_{\rm Lab} = 90 A$ MeV to  $\rm E_{\rm cm} = 200 A$ GeV and also, showed disagreement with the available data. In Ref~~\cite{Petersen:2009vx}, the excitation function of $v_2$ was examined in the range of GSI-SIS to CERN-SPS energies using UrQMD with HG EoS within hybrid approach and other harmonics such as, $v_{2}$ and $v_{3}$ are studied with Chiral EoS in Au--Au systems between $\sqrt{s_{NN}}$ $=$ 5--200 GeV~\cite{Auvinen:2013sba}. The collision energy dependence of $v_1$ is tested using the hybrid model for nuclear reactions between $\sqrt{s_{NN}}= 3 -20$ GeV~\cite{Steinheimer:2014pfa}. In our previous work~\cite{Rode:2019pey}, study on nuclear equations of state dependence of anisotropic flow was performed using hybrid UrQMD model within 6A--25A GeV with HG and chiral EoS. All these results seem to suggest the quantitative applicability of this model to real scenario has some limitations. However, in this paper, we qualitatively aim to understand the effect of various nuclear equations of state on the flow harmonics and hence gain some insights about the dynamics leading to their development at various beam energies ranging from 1A--158A GeV.

Besides anisotropic flow, we also attempt to study the EoS dependence of particle production in non-central collisions. The particle ratios of various species are examined for this purpose. We also look at the net-proton rapidity distributions. The structure of the net-proton rapidity spectra at the mid rapidity is expected to be sensitive to the underlying EoS of the nuclear fireball. In central collisions, adequate studies have been performed in this direction. In Refs.~\cite{Ivanov:2010cu,Ivanov:2012bh,Ivanov:2013mxa,Ivanov:2015vna,Ivanov:2016xev}, the authors have quantified the structure of net proton rapidity distribution at mid rapidity in central collisions, in terms of reduced curvature. It was studied as a function of beam energy and compared with predictions incorporating various possible scenarios of fireball expansion. In the present article we extend these studies to the mid central collisions.

This article is arranged in the following order. In section II, basic principle of UrQMD model and its different variants are briefly introduced. The obtained results on anisotropic flow coefficients and particle production properties over a very wide range of colliding energies are presented in section III. Finally we summarize the results in section IV.

\section*{II Model description}
For detailed description of the Ultra-relativistic Quantum Molecular Dynamics (UrQMD) model, the reader is referred to Refs.~\cite{Bass:1998ca,Bleicher:1999xi,Petersen:2008dd}. The purpose of the UrQMD model is to simulate high energy nucleus-nucleus collisions. The initializations of the target and projectile nuclei in co-ordinate and momentum space are done with the help of Woods-Saxon profile and Fermi gas model, respectively. Together with the various experimental inputs such as cross-sections, decay widths, the collisions in the model are narrated in terms of interactions among resonances, hadrons and their excited states at low energies and in terms of excitations of color strings with their subsequent fragmentation into hadrons at higher energies~\cite{Bleicher:1999xi}.  The propagation of hadrons is taken place on straight line trajectories amid subsequent collisions. 

In the hybrid version of the UrQMD, the ideal (3+1)d relativistic fluid dynamical evolution using SHASTA~\cite{Rischke:1995ir,Rischke:1995mt} algorithm is combined with pure transport approach for a better modeling of the intermediate hot and dense stages of the collision. The calculation of initial state of the hydrodynamical evolution is crucial to account for non-equilibrium nature of the early stage, moreover, this also incorporates event-by-event fluctuations of the initial states. The hydrodynamical evolution is commenced upon crossing of the two Lorentz-contracted nuclei~\cite{Petersen:2008dd}. This choice of the initial time ensures that all initial baryon-baryon scatterings and consequent energy deposition have taken place and represents the  lower limit of the time scale for thermalization. Thereafter, the mapping of particles which are treated as "point-like" in the initial stage, to hydrodynamic grid is performed while the spectators are propagated in the cascade. This is immediately followed by the hydrodynamical evolution for which equation of state (EoS) serves as one of the important inputs. After dropping of local energy density $\epsilon$ below five times the ground state energy density $\epsilon_0$~\cite{Petersen:2008dd}, the hydrodynamical evolution ceases and the hadronization is performed by mapping the hydrodynamical fields to the hadrons using Cooper-Frye prescription~\cite{Cooper:1974mv}. The model offers two different freeze-out criteria~\cite{Santini:2011rq}. In the isochronous freeze-out (IF) scenario, all hydrodynamic cells are mapped onto particles on same time, provided the energy density drops below the critical value in all cells. Alternatively in gradual freeze-out (GF) scenario, 0.2 fm thick transverse slices are particlized when energy density in all cells of each individual slice drops below the critical value. Employment of gradual transition leads to a rapidity independent transition temperature without artifical time dilation effects. In our calculations we use GF scenario~\cite{Steinheimer:2009nn}.  Authors of Ref.~\cite{Petersen:2009mz} have seen that such freeze-out conditions provides best description of the data for mean transverse mass excitation function. Thereafter, the hadrons are evolved through rescatterings and decays until the decoupling of the system.

In hydro mode, there are several available EoS that can be employed. One of them is Hadron Gas (HG) EoS~\cite{Zschiesche:2002zr} which has similar underlying degrees of freedom as pure transport approach. It consists of non-interacting gas of hadrons described by grand canonical ensemble and does not incorporate any type of phase transition. This gives an excellent opportunity to compare the hydrodynamical and pure transport approach on equal footings. 

\begin{figure*}
\begin{center}
\includegraphics[scale=0.3]{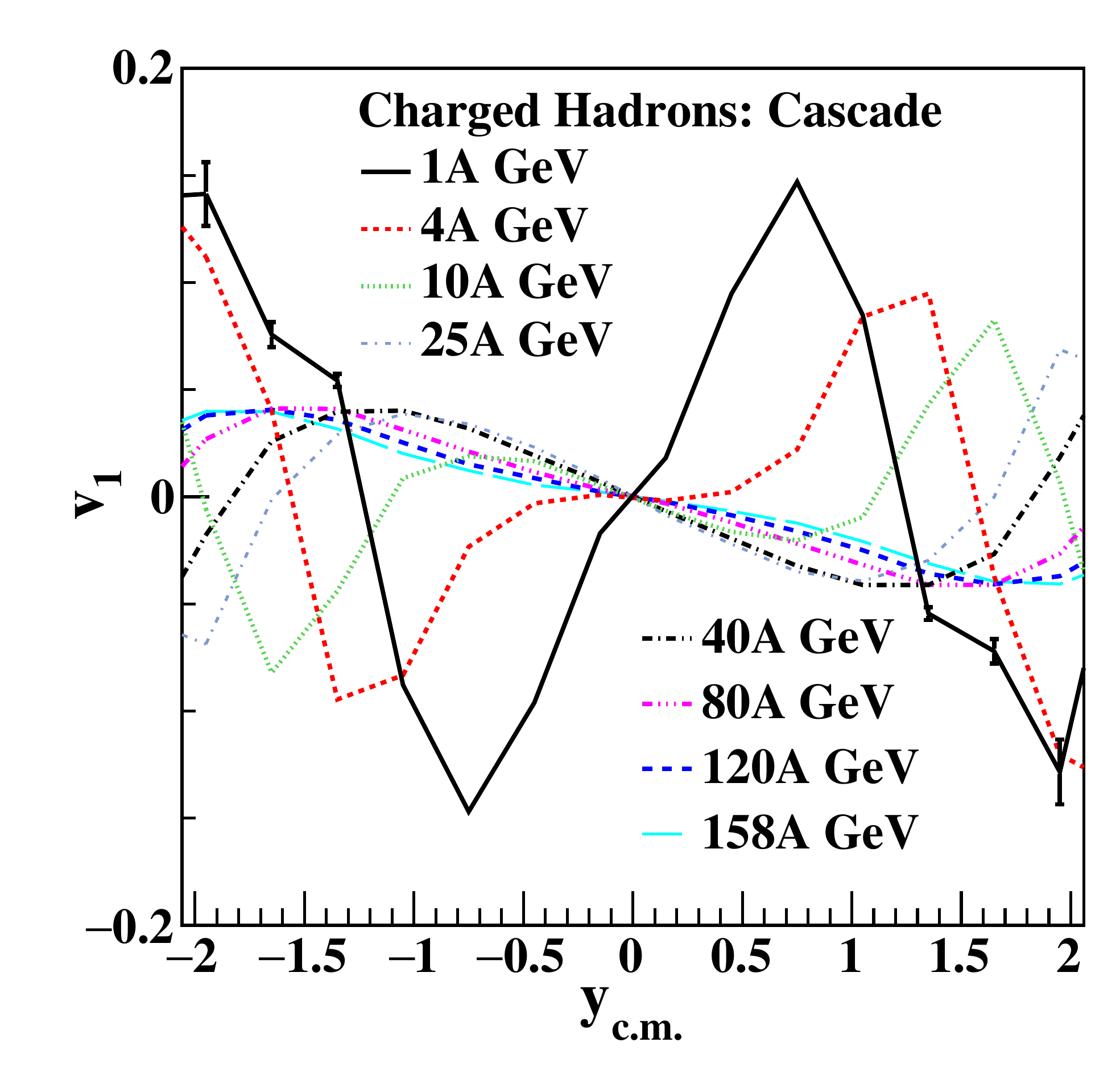}
\includegraphics[scale=0.3]{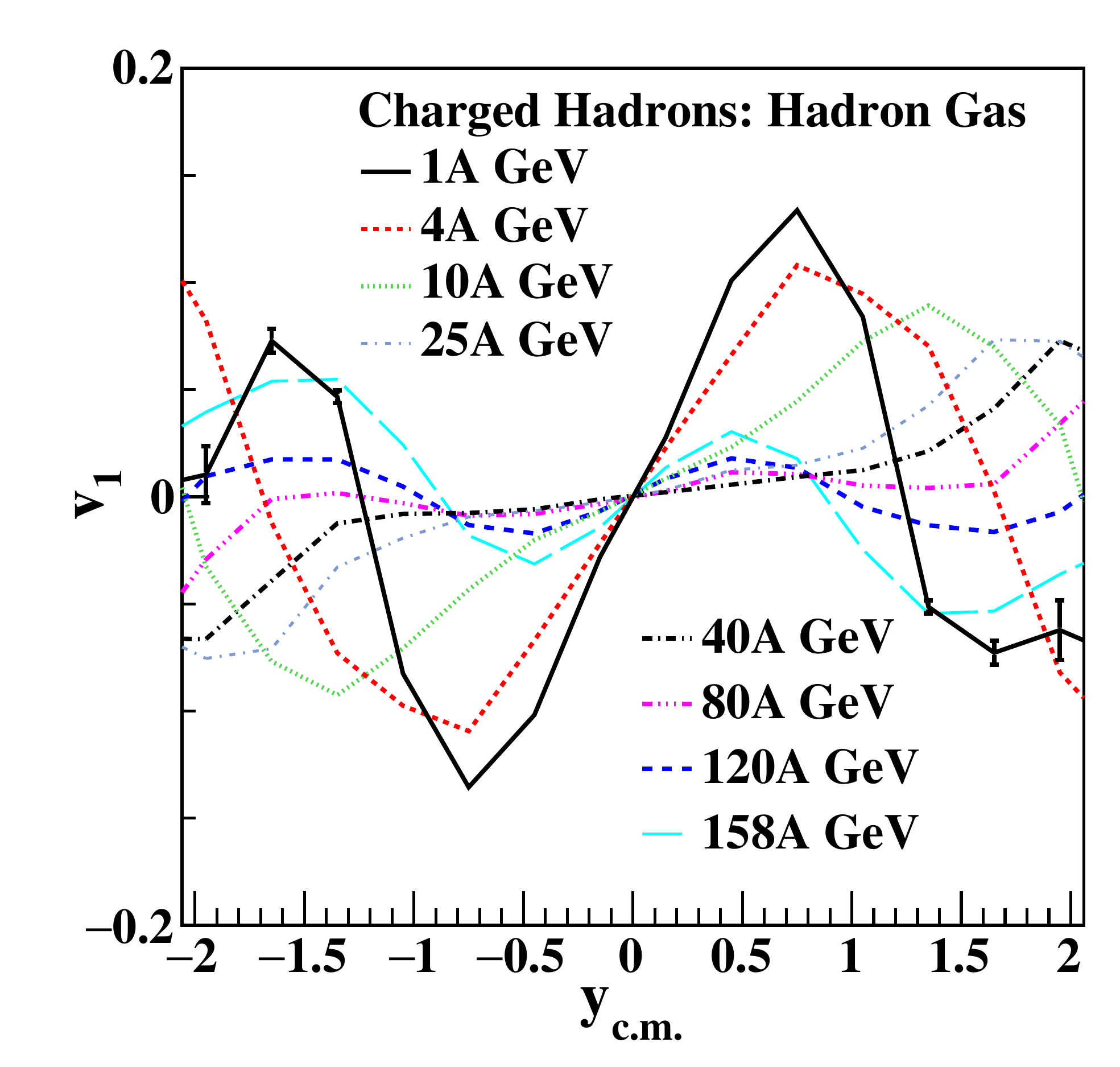}\\
\includegraphics[scale=0.3]{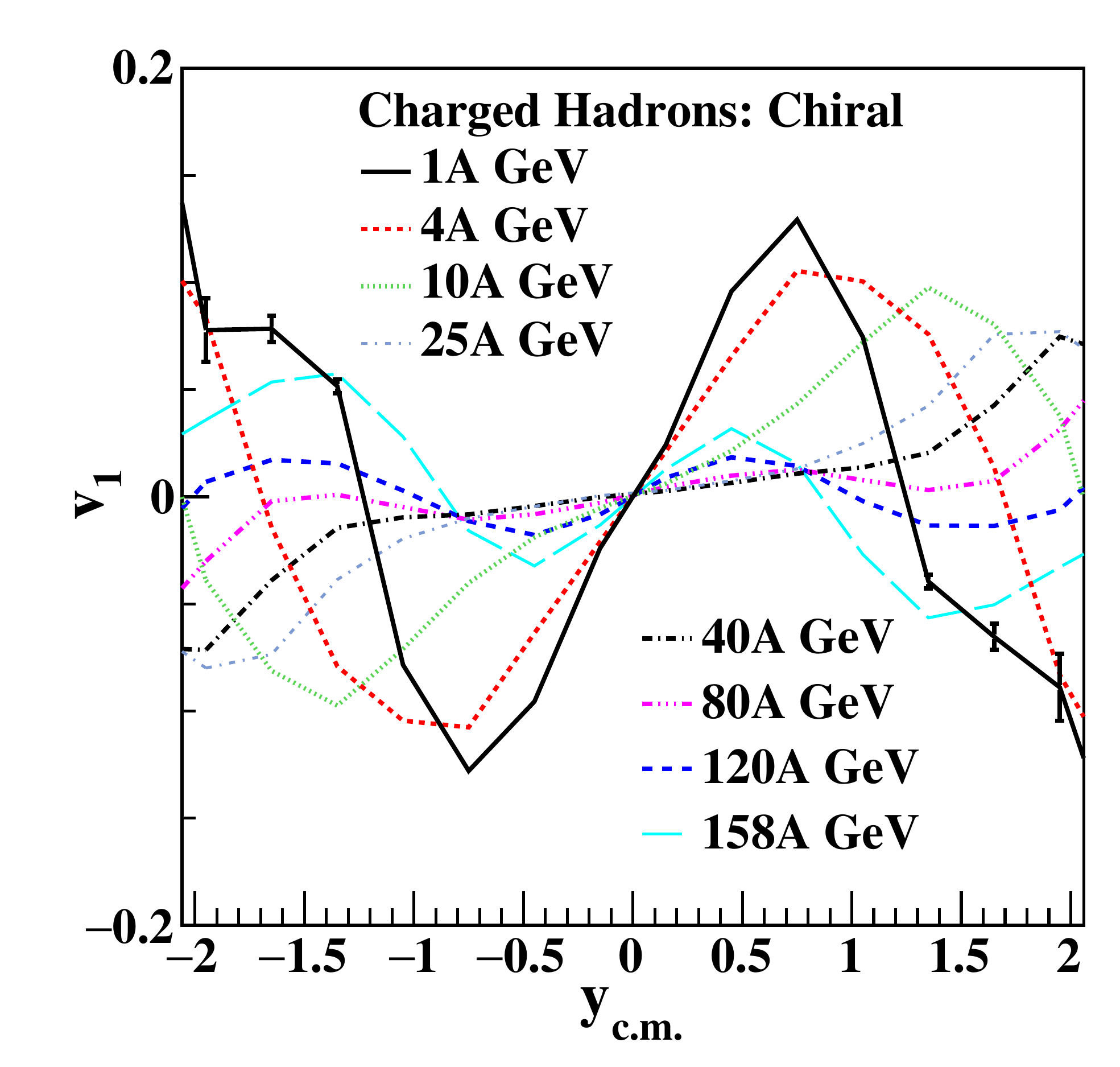}
\includegraphics[scale=0.3]{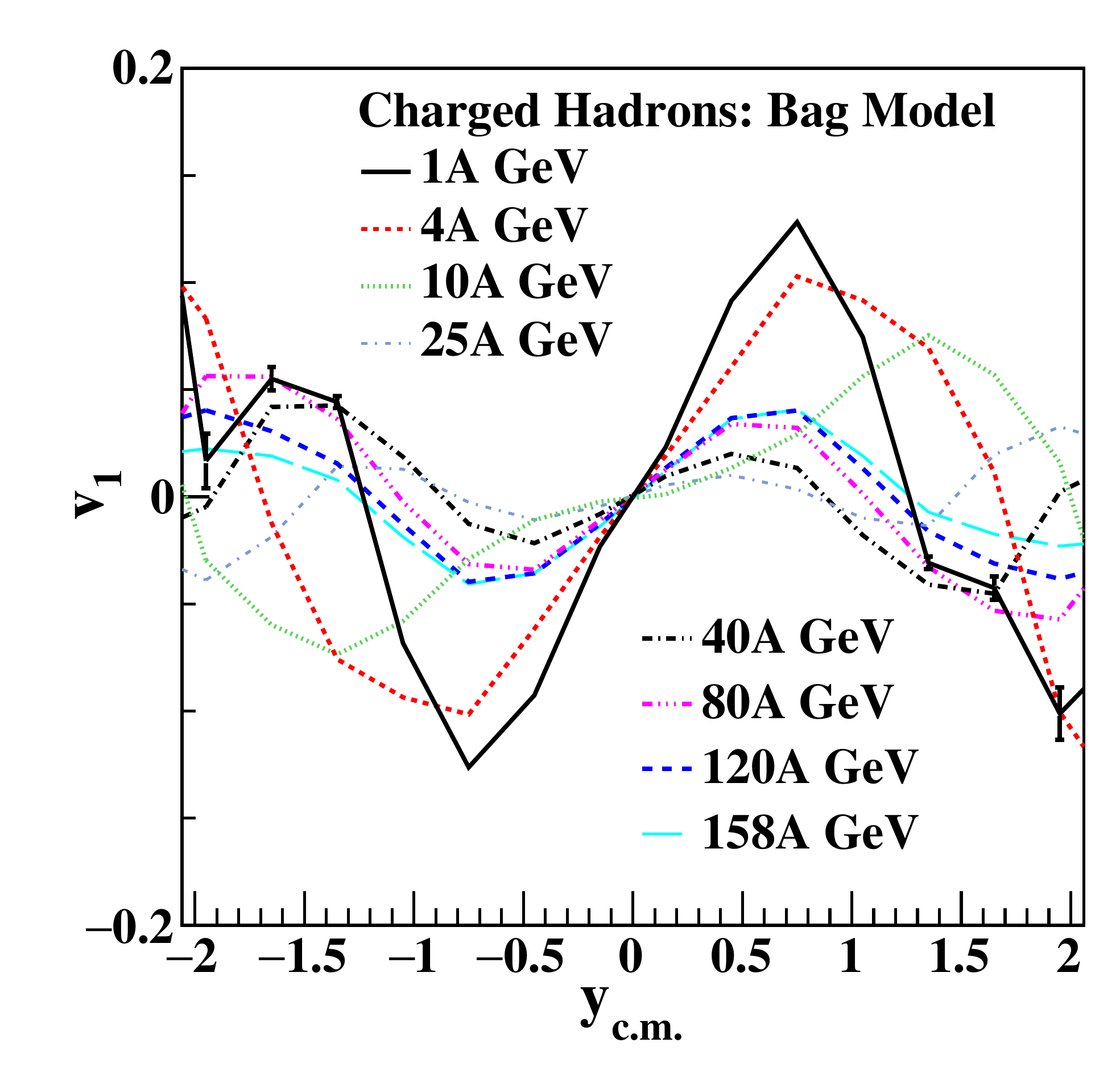}
\end{center}
\caption{Directed flow of charged hadrons as a function of rapidity at different beam energies for different configurations of UrQMD for non-central (b = 5-9 fm corresponds to approximately 10-40$\%$ central) Au-Au collisions.}
\label{fig1}
\end{figure*}

\begin{figure*}
\begin{center}
\includegraphics[scale=0.3]{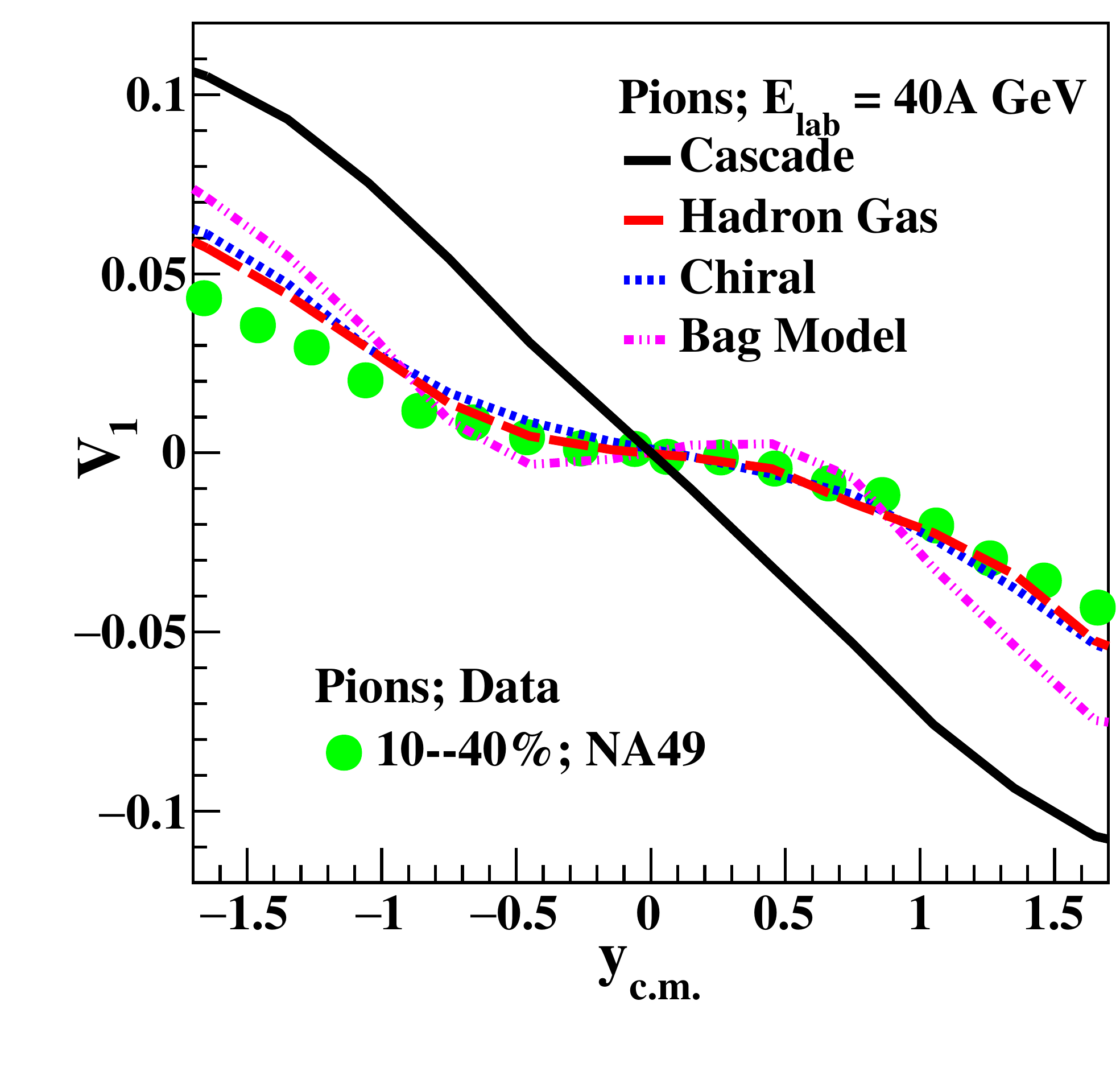}
\includegraphics[scale=0.3]{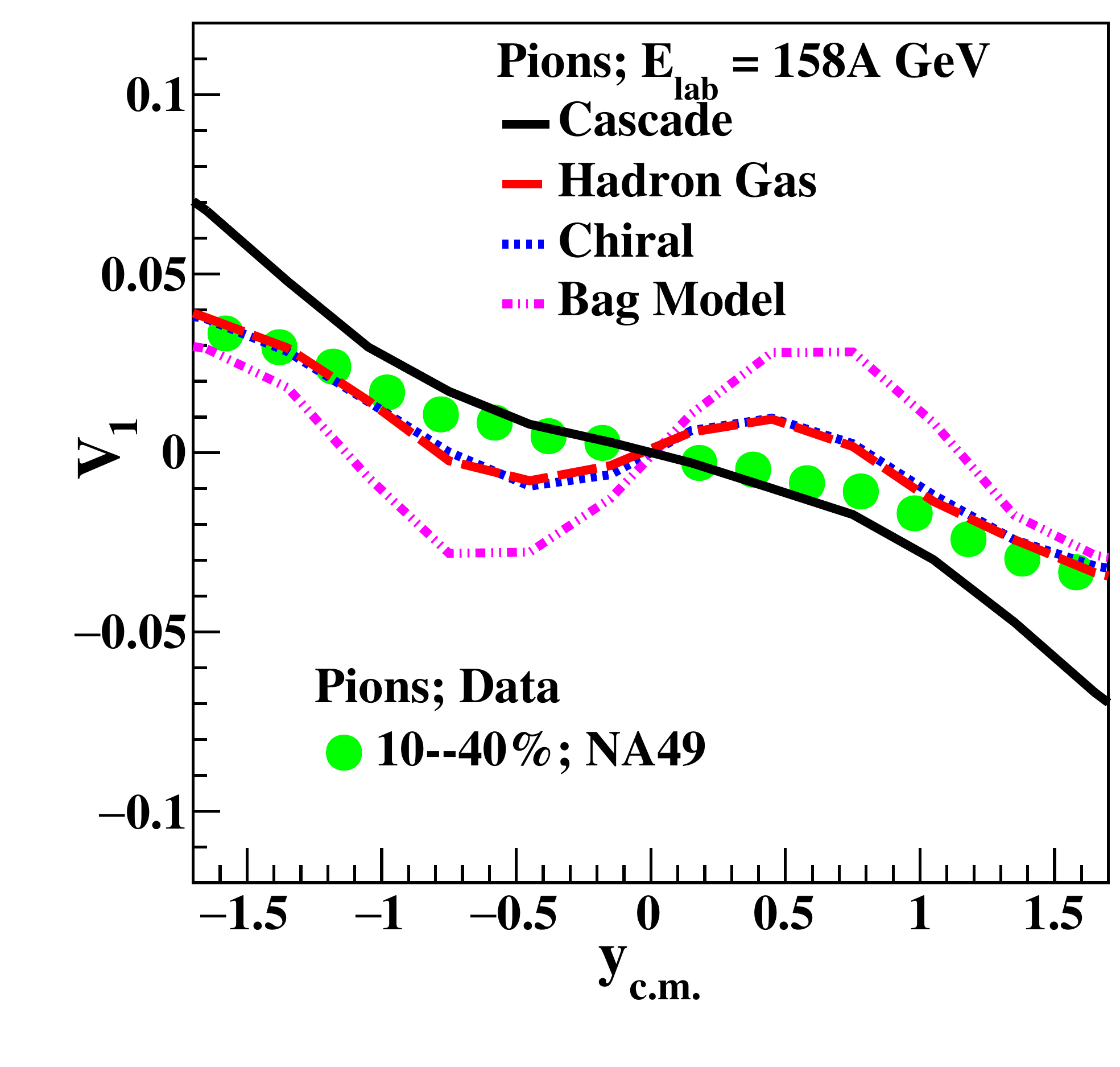}\\
\includegraphics[scale=0.3]{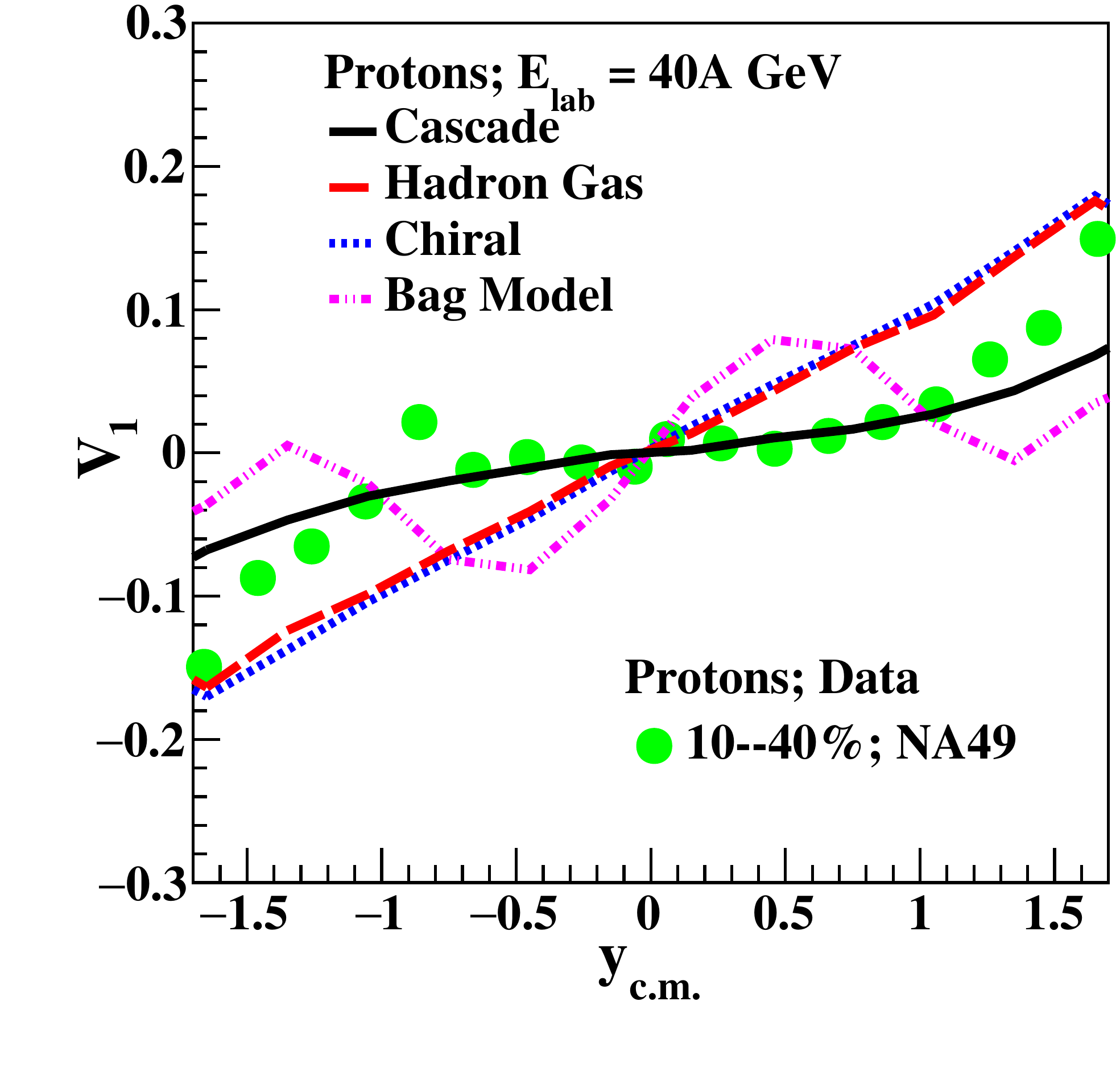}
\includegraphics[scale=0.3]{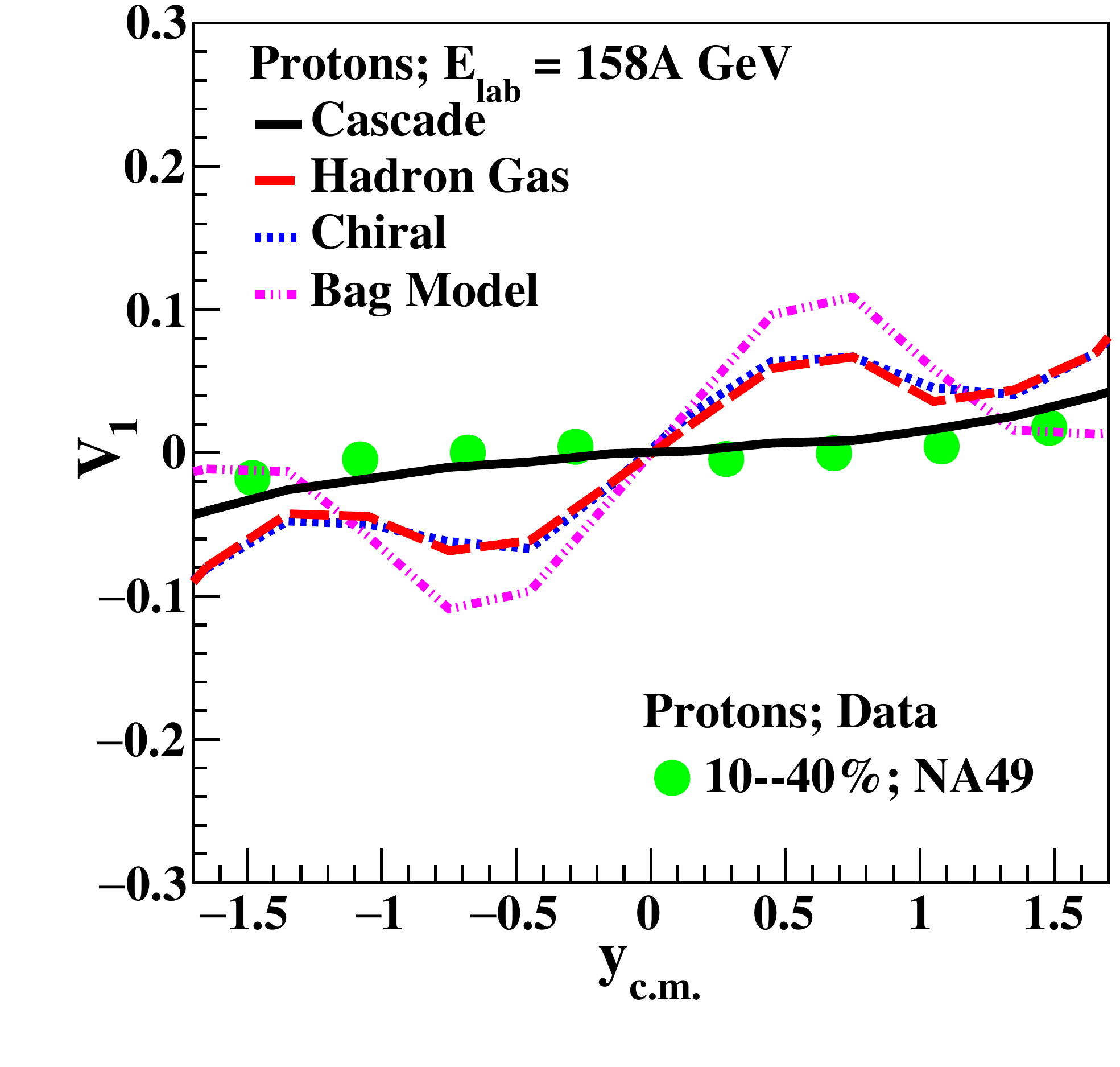}
\end{center}
\caption{Comparison of directed flow of pions and protons as a function of rapidity for different configurations of UrQMD with measured directed flow for $p_{\rm T}$ $<$ 2 GeV/$c$ for non-central (b = 5-9 fm corresponds to approximately 10-40$\%$ central) Au-Au collisions with NA49 experimental measurements~\cite{Alt:2003ab} at 40A and 158A GeV in Pb-Pb collisions.}
\label{fig2}
\end{figure*}

\begin{figure*}
\begin{center}
\includegraphics[scale=0.3]{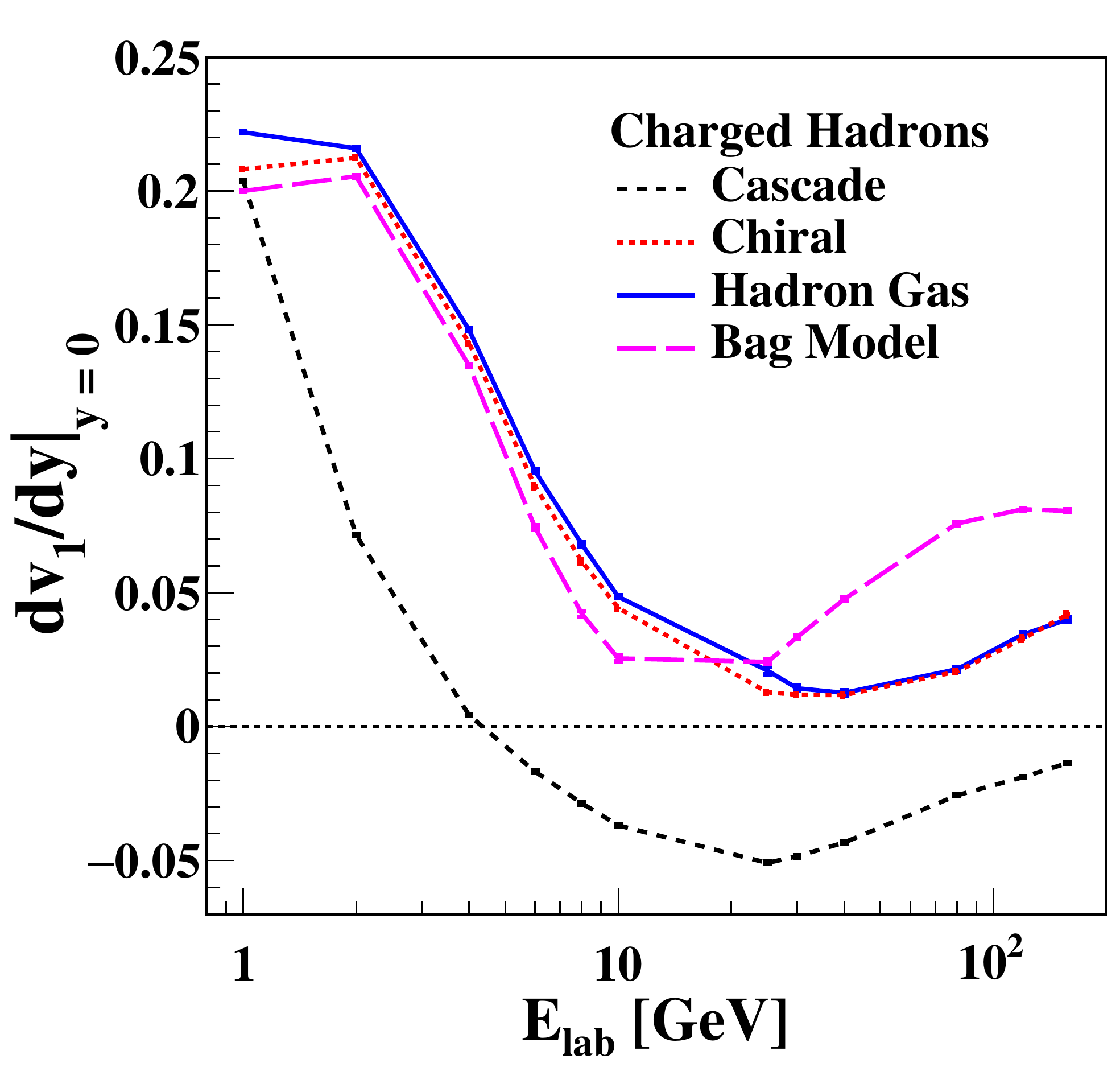}
\includegraphics[scale=0.3]{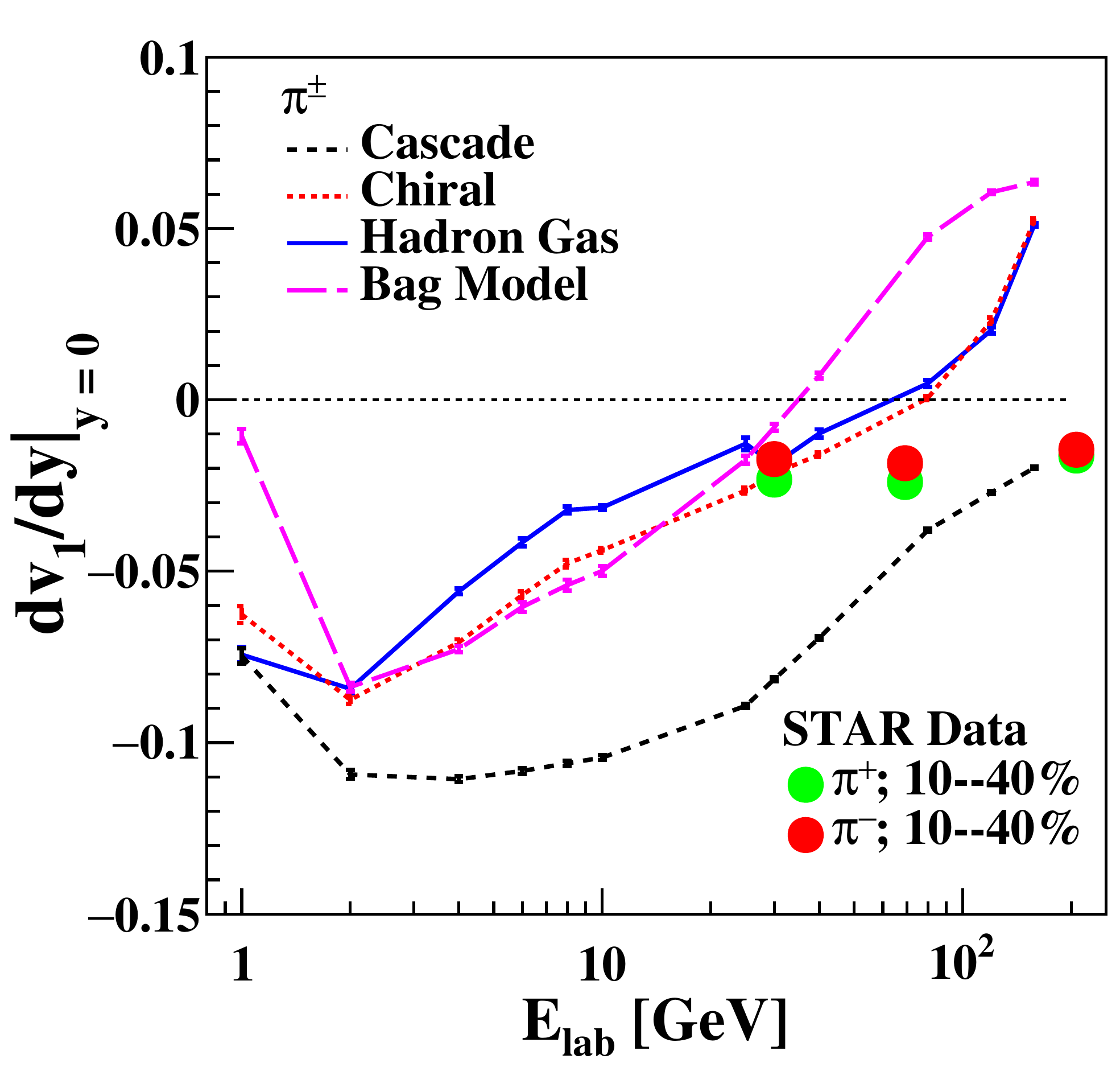}\\
\includegraphics[scale=0.3]{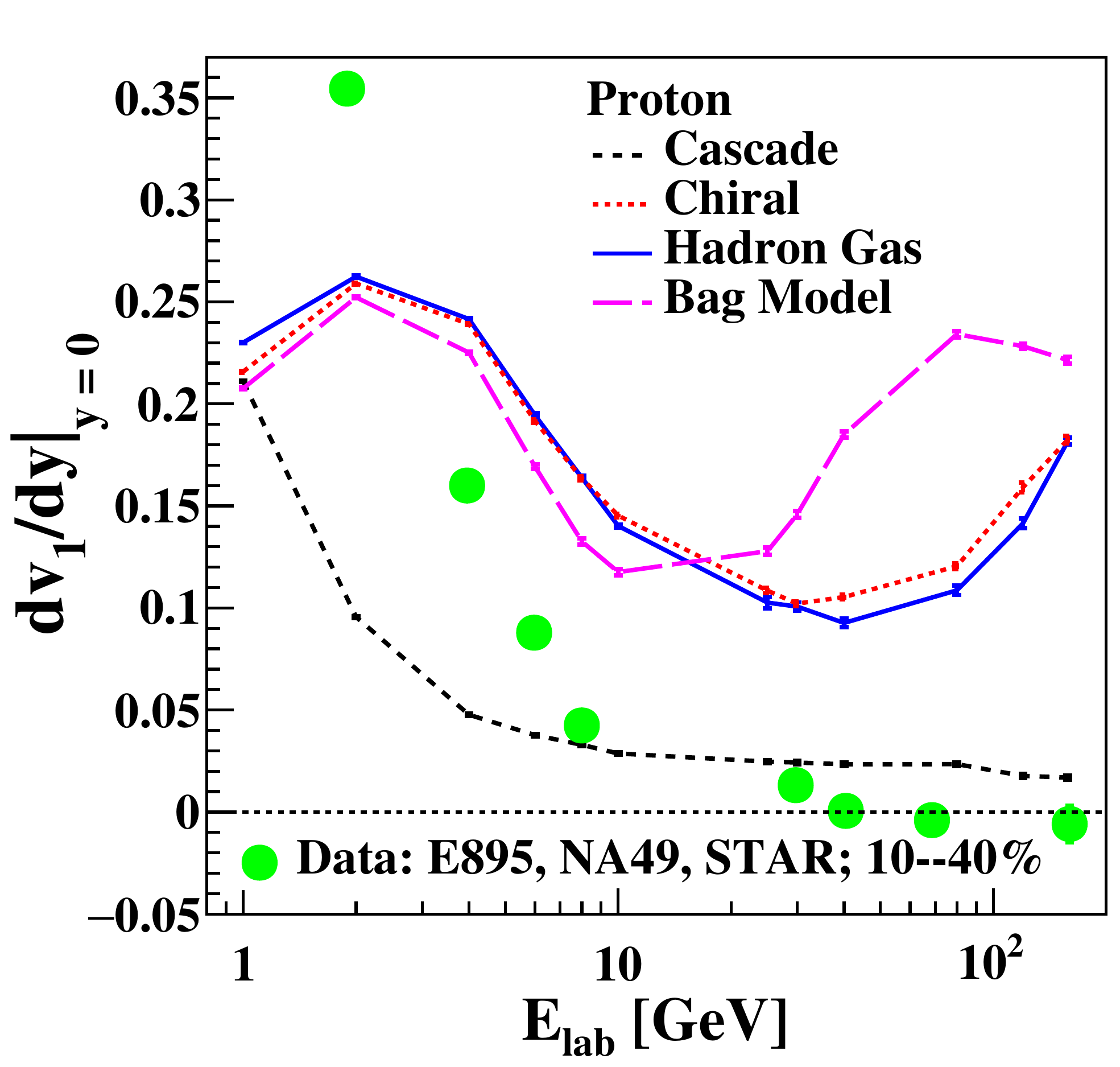}
\includegraphics[scale=0.3]{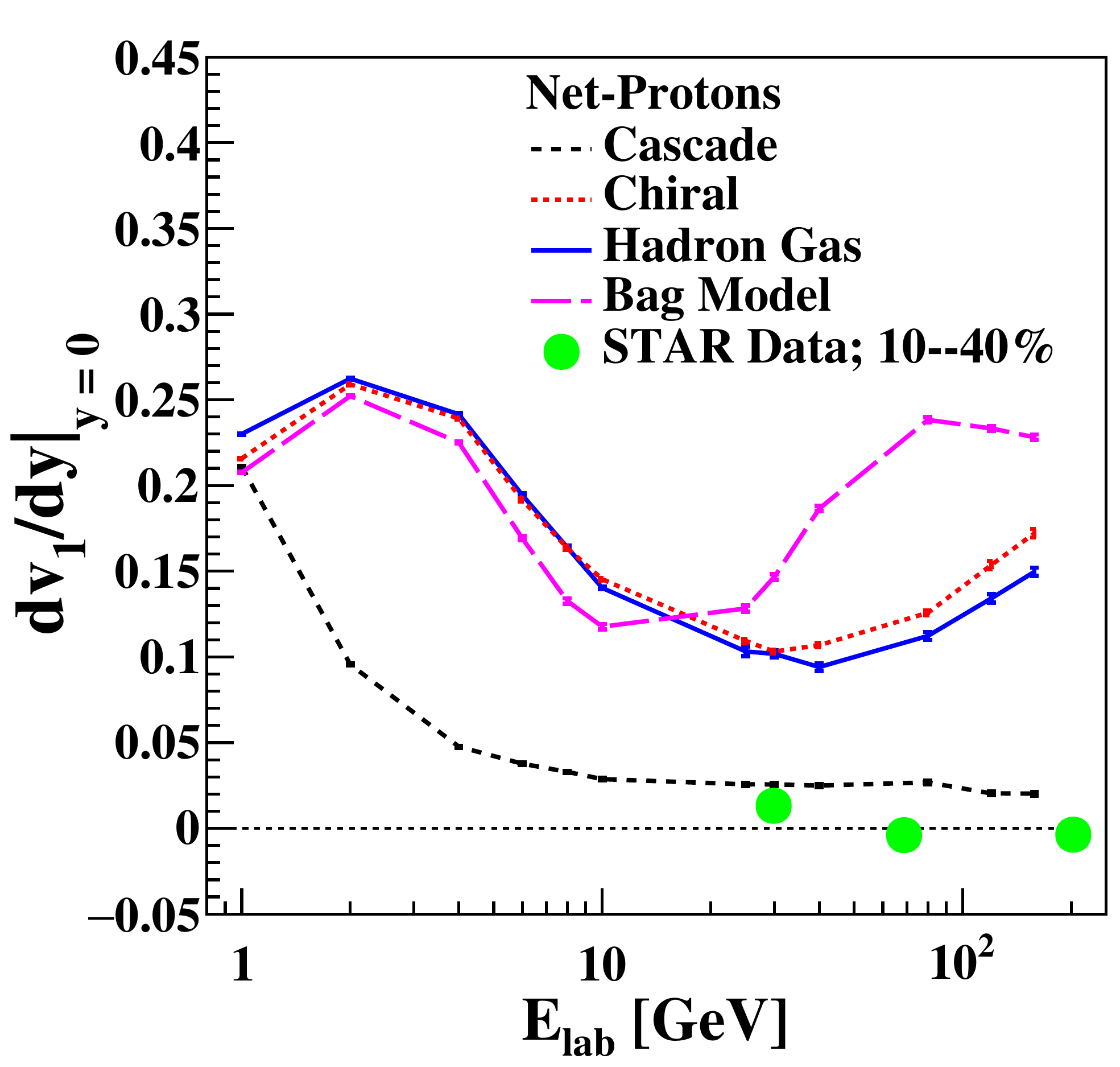}
\end{center}
\caption{Slope of the directed flow of charged hadrons, pions, protons and net-protons as a function of beam energy at midrapidity for different configurations of UrQMD for non-central (b = 5-9 fm corresponds to approximately 10-40$\%$ central) Au-Au collisions with E895~\cite{Liu:2000am} and STAR~\cite{Adamczyk:2014ipa} experimental measurements in Au-Au collisions and with NA49~\cite{Alt:2003ab} experimental measurements in Pb-Pb collisions.}
\label{fig3}
\end{figure*}

\begin{figure*}
\begin{center}
\includegraphics[scale=0.3]{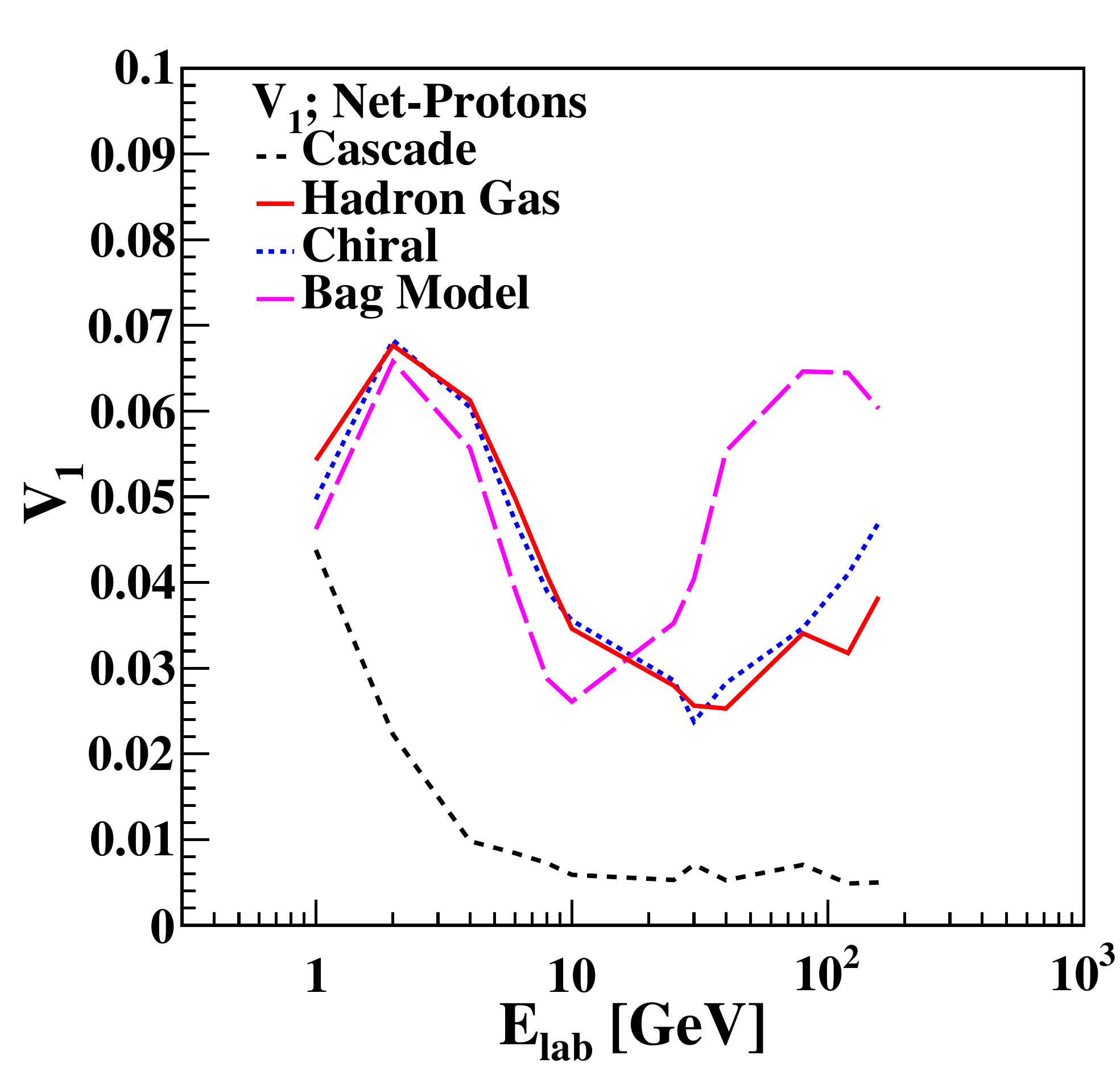}
\includegraphics[scale=0.3]{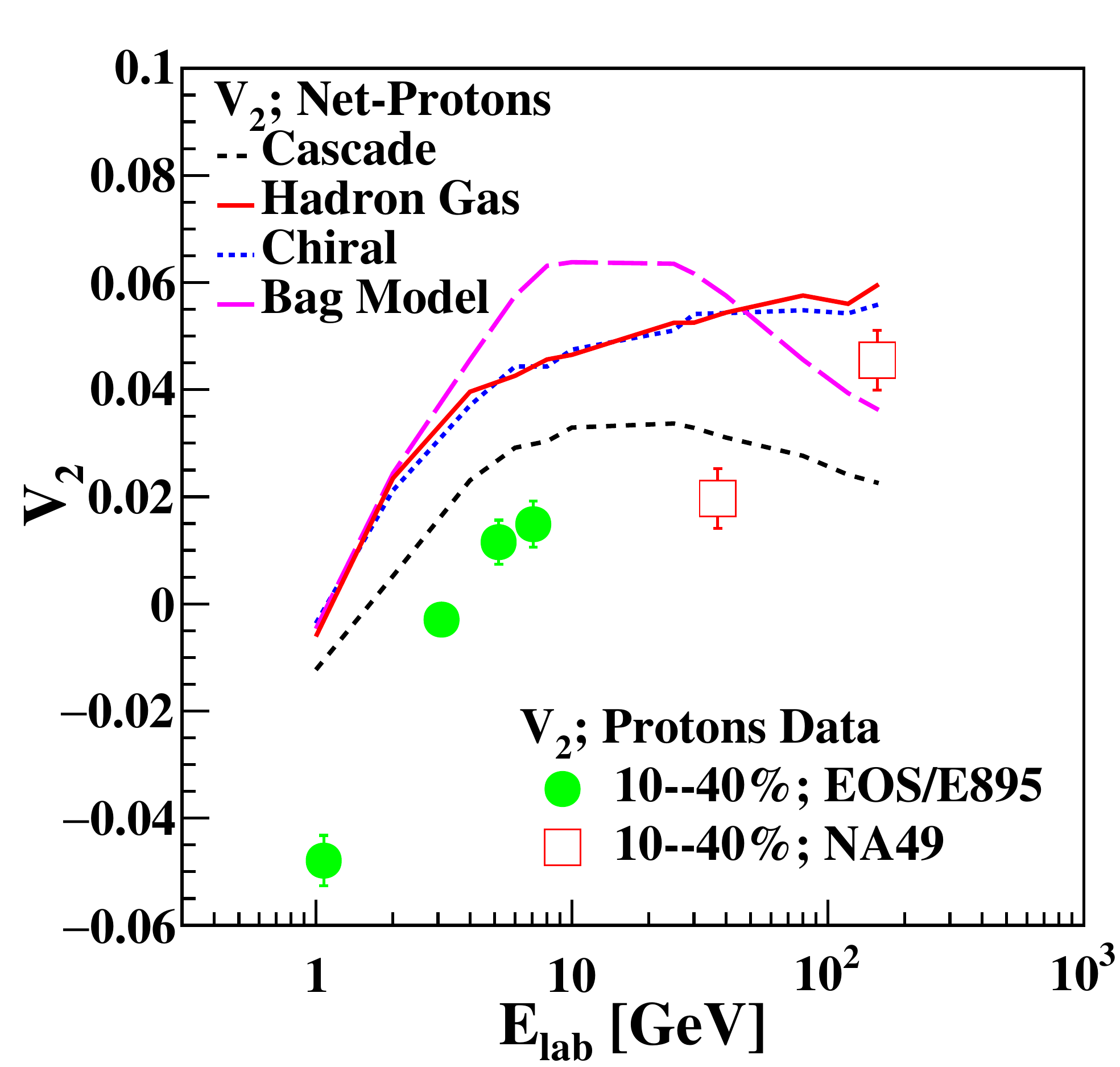}
\end{center}
\caption{$p_{\rm T}$ integrated directed $(v_{1})$ (left) and elliptic $(v_{2})$ (right) flow of net-protons as a function of beam energy at midrapidity (0 $<$ $y_{c.m.}$ $<$ 0.5) for different configurations of UrQMD for non-central (b = 5-9 fm corresponds to approximately 10-40$\%$ central) Au-Au collisions.  In the right plot, $v_{2}$ of protons for $p_{\rm T}$ $<$ 2 GeV/$c$ is compared with available E895 and NA49 experimental measurements~\cite{Pinkenburg:1999ya,Alt:2003ab} in investigated beam energy range in Au-Au and Pb-Pb collisions, respectively.}
\label{fig4}
\end{figure*}

The other possible choice includes the BAG model EoS~\cite{Rischke:1995mt}. It has an in built first order deconfinement phase transition anticipated at finite baryon densities. In this EoS, an improved version of $\sigma-\omega-$ model with realistic effective nucleon mass and ground state incompressibility values is employed in case of hadronic matter whereas, standard MIT bag model~\cite{Chodos:1974je} is recruited for the QGP phase. During the transition, both these phases are matched with the help of Gibbs' conditions for phase equilibrium~\cite{Rischke:1995mt}.

\begin{figure*}
\begin{center}
\includegraphics[scale=0.3]{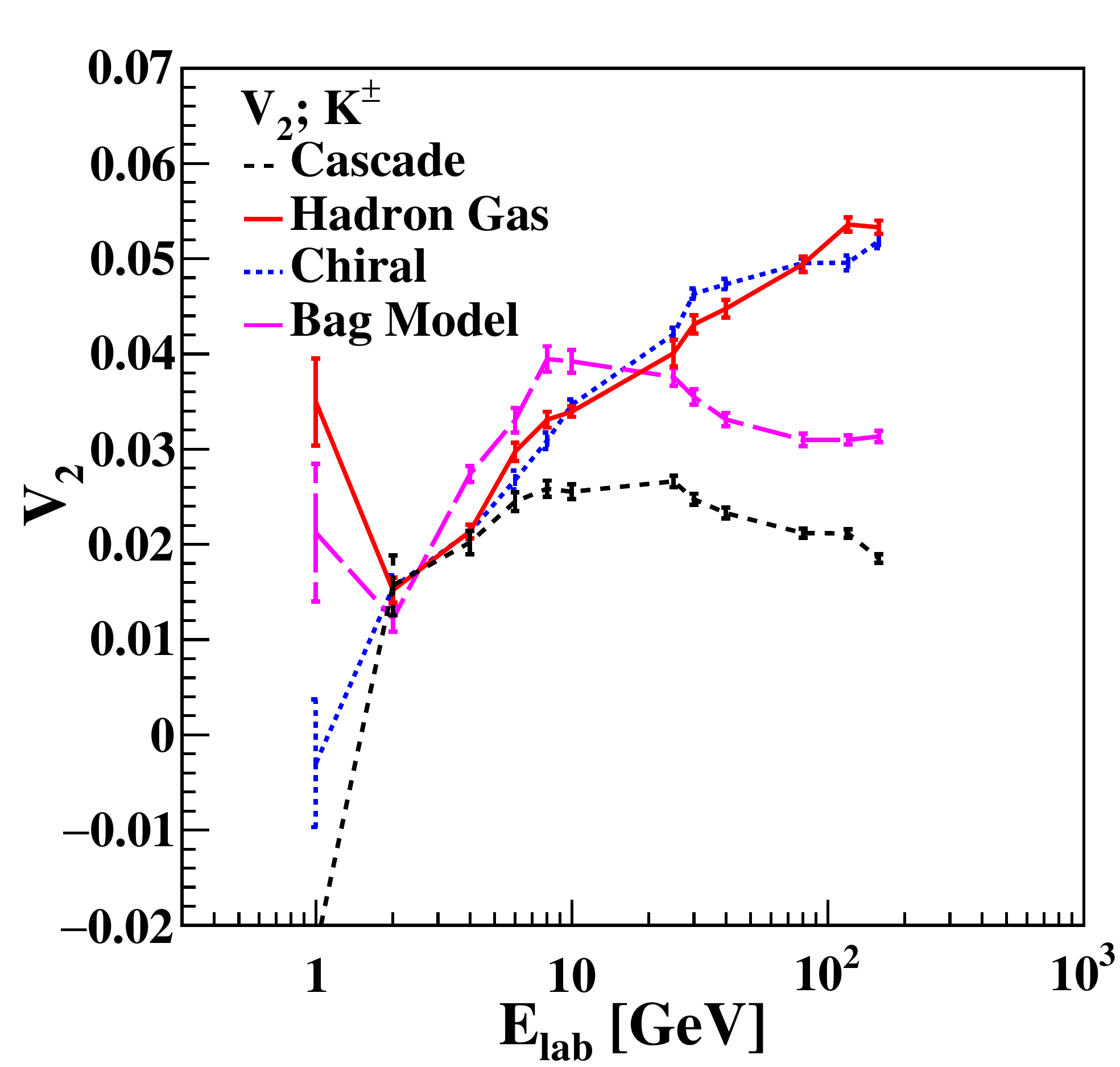}
\includegraphics[scale=0.3]{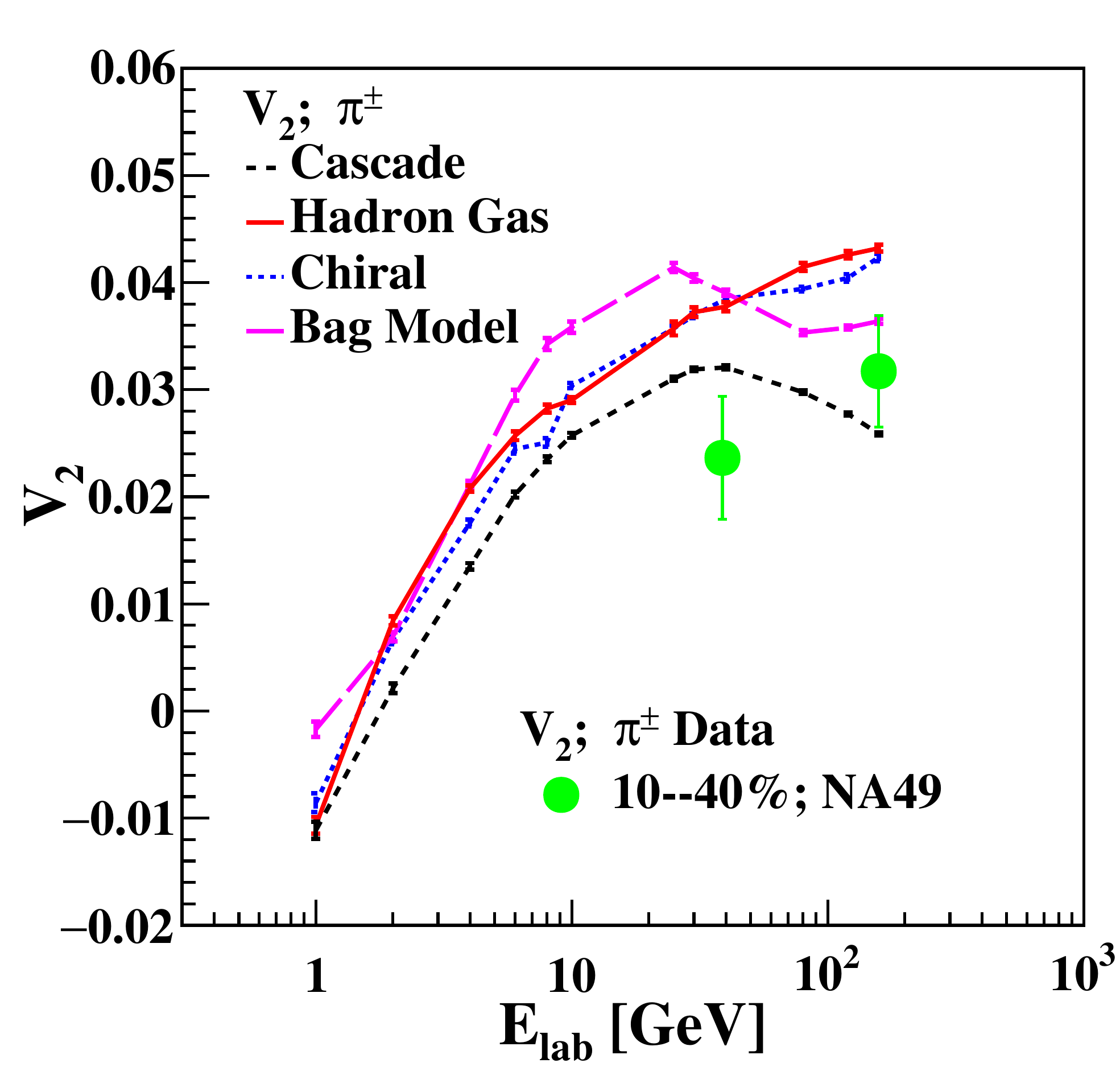}
\end{center}
\caption{$p_{\rm T}$ integrated elliptic flow of kaons and pions as a function of beam energy at midrapidity (-0.5 $<$ $y_{c.m.}$ $<$ 0.5) for different configurations of UrQMD for non-central (b = 5-9 fm corresponds to approximately 10-40$\%$ central) Au-Au collisions. In the right plot, $v_{2}$ of pions for $p_{\rm T}$ $<$ 2 GeV/$c$ is compared with available NA49 experimental measurements~\cite{Alt:2003ab} in investigated beam energy range in Pb-Pb collisions.}
\label{fig5}
\end{figure*}

\begin{figure*}
\begin{center}
\includegraphics[scale=0.5]{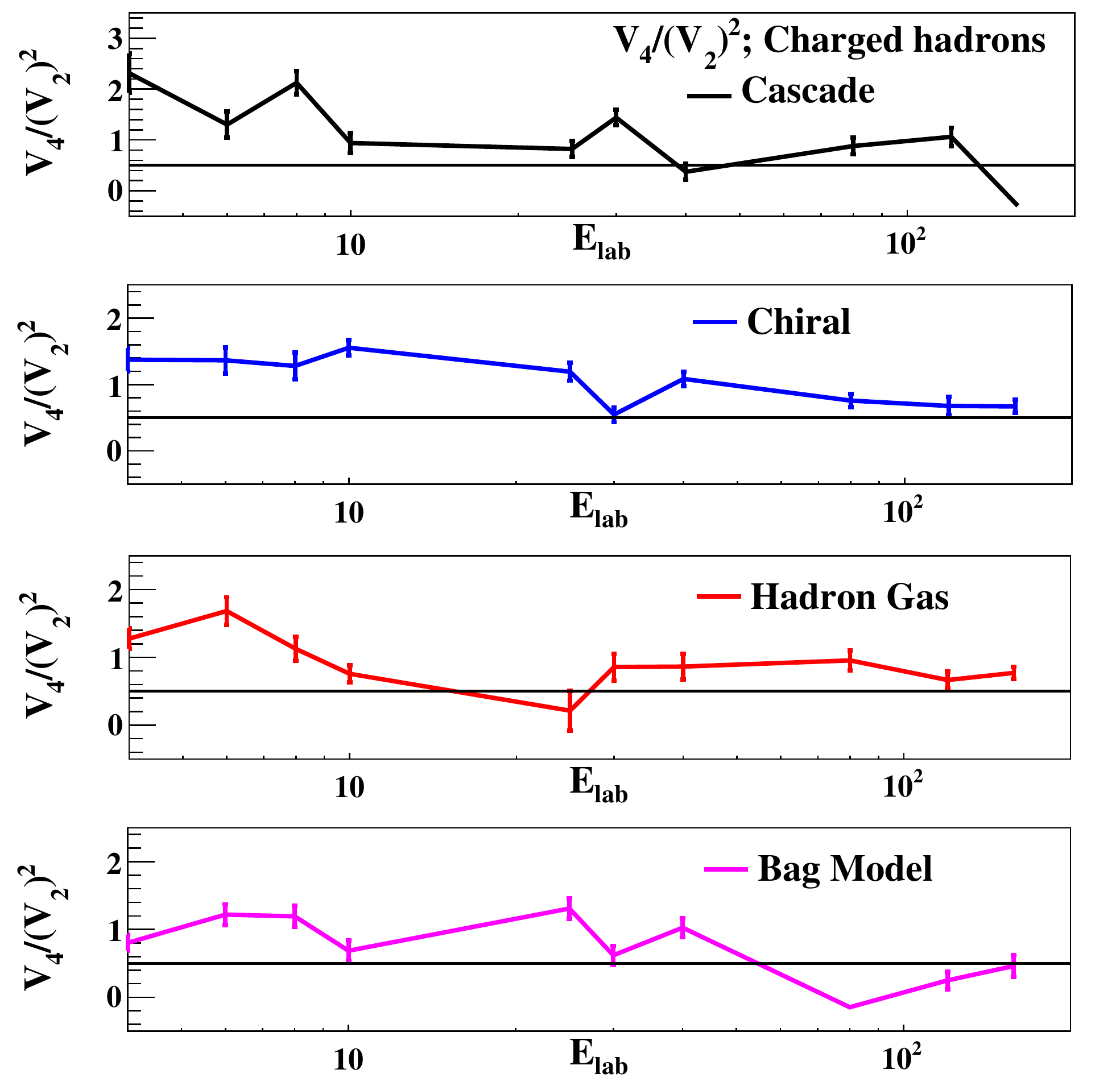}
\end{center}
\caption{$V_{4}/(V_{2})^{2}$ of charged hadrons as a function of beam energy at midrapidity (-0.5 $<$ $y_{c.m.}$ $<$ 0.5) for different configurations of UrQMD for non-central (b = 5-9 fm corresponds to approximately 10-40$\%$ central) Au-Au collisions. The horizontal line at 0.5 denote the ideal fluid dynamic limit.}
\label{figV4}
\end{figure*}

Moving on, there is another available EoS named Chiral + deconfinement EoS~\cite{Steinheimer:2011ea} employed in this investigation. Both chiral as well as deconfinement phase transitions are included in this EoS while the latter is a continuous cross over all finite net baryon densities ($\mu_{B}$). The chiral phase transition is administrated by hadronic interactions whereas, deconfinement transition via quarks and Polyakov potential. The partonic degrees of freedom only show up at higher temperatures where hadrons disappear. At vanishing $\mu_B$ this EoS matches well with the lattice QCD simulations. Independent use of three different EoS within the hybrid UrQMD model enables us to compare three distinct fireball evolution scenarios over the entire enrgy domain investigated in this work.
\begin{figure*}[h!]
\begin{center}
\includegraphics[scale=0.30]{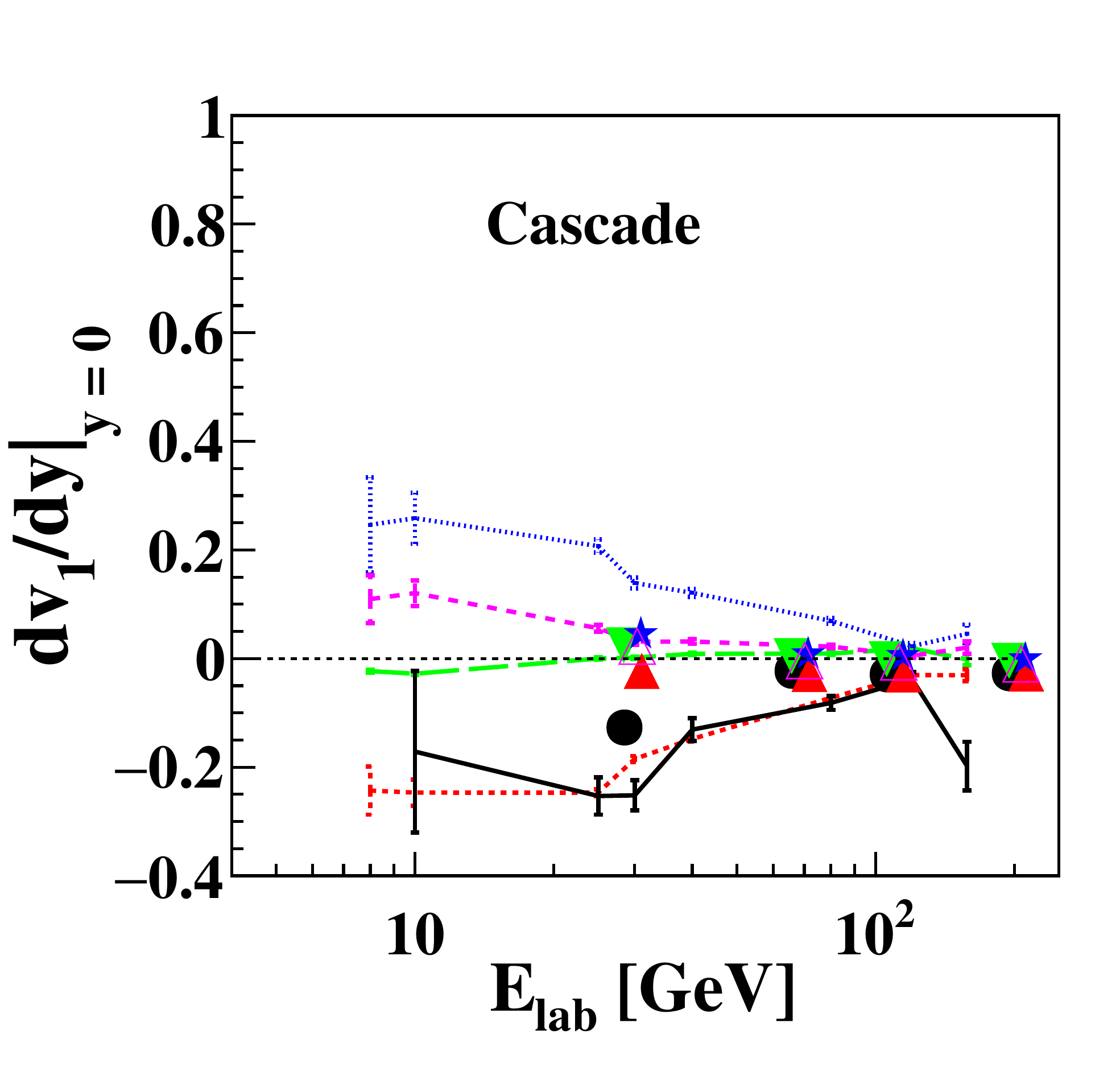}
\includegraphics[scale=0.30]{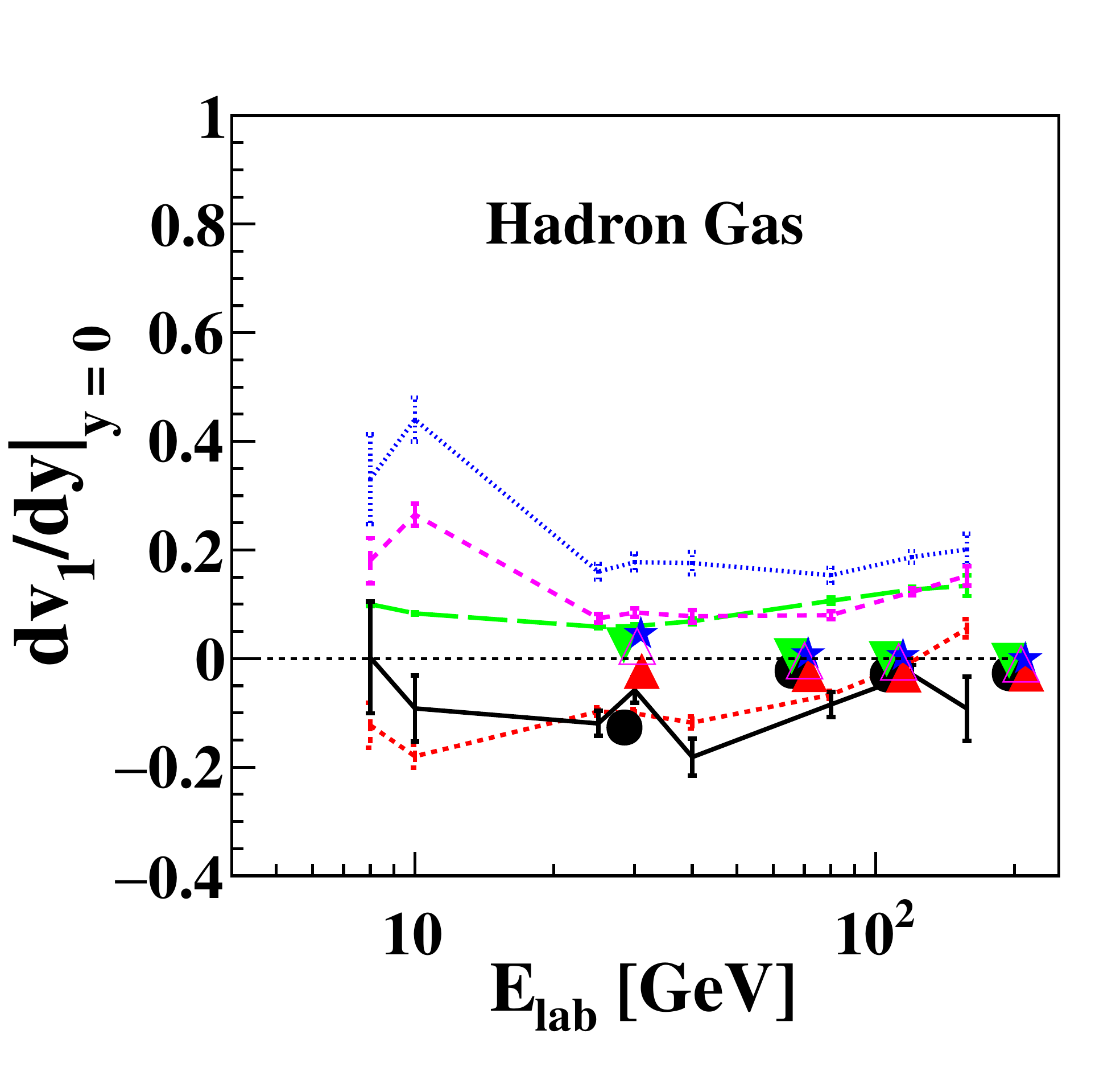}
\includegraphics[scale=0.28]{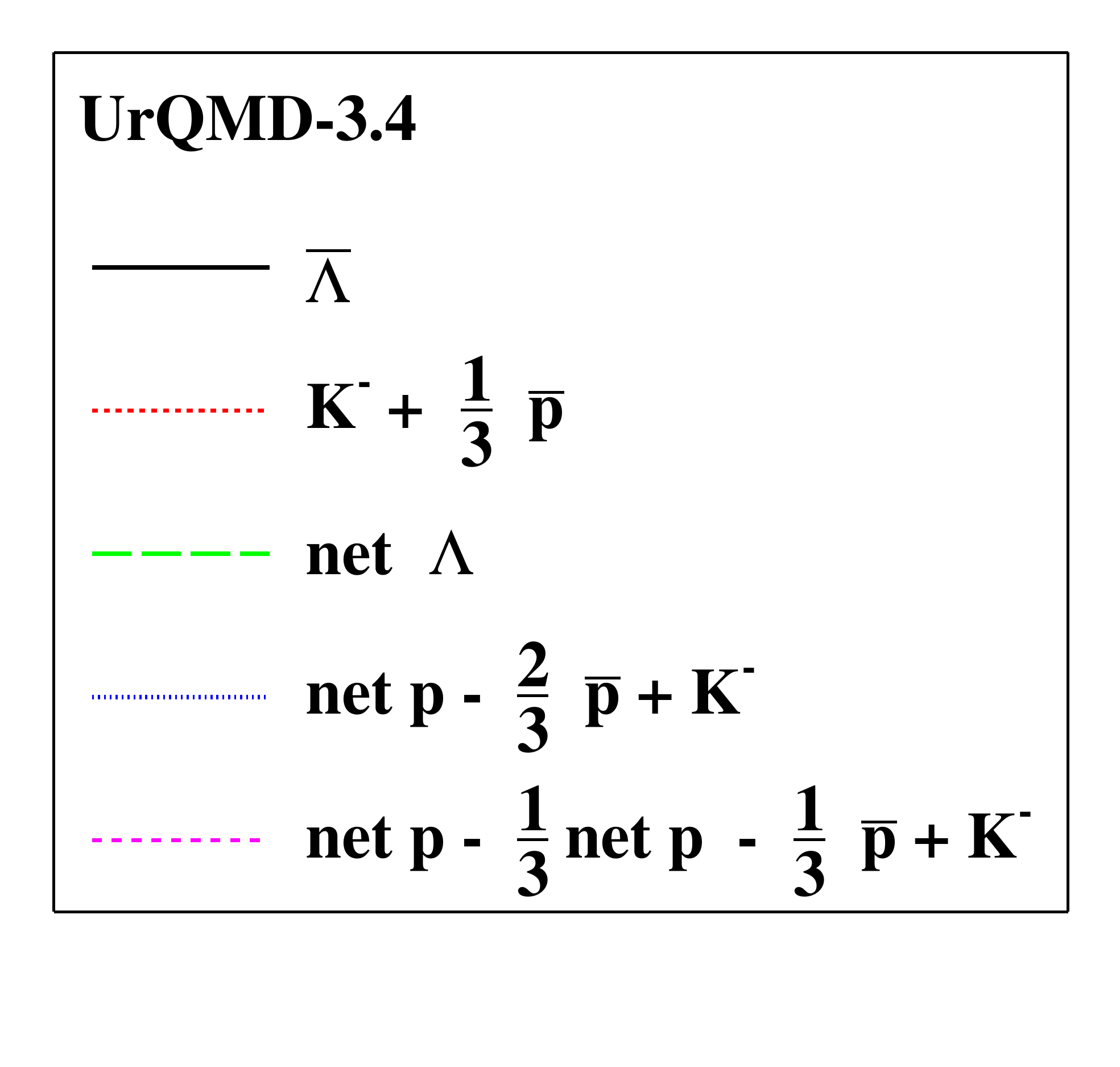}\\
\includegraphics[scale=0.30]{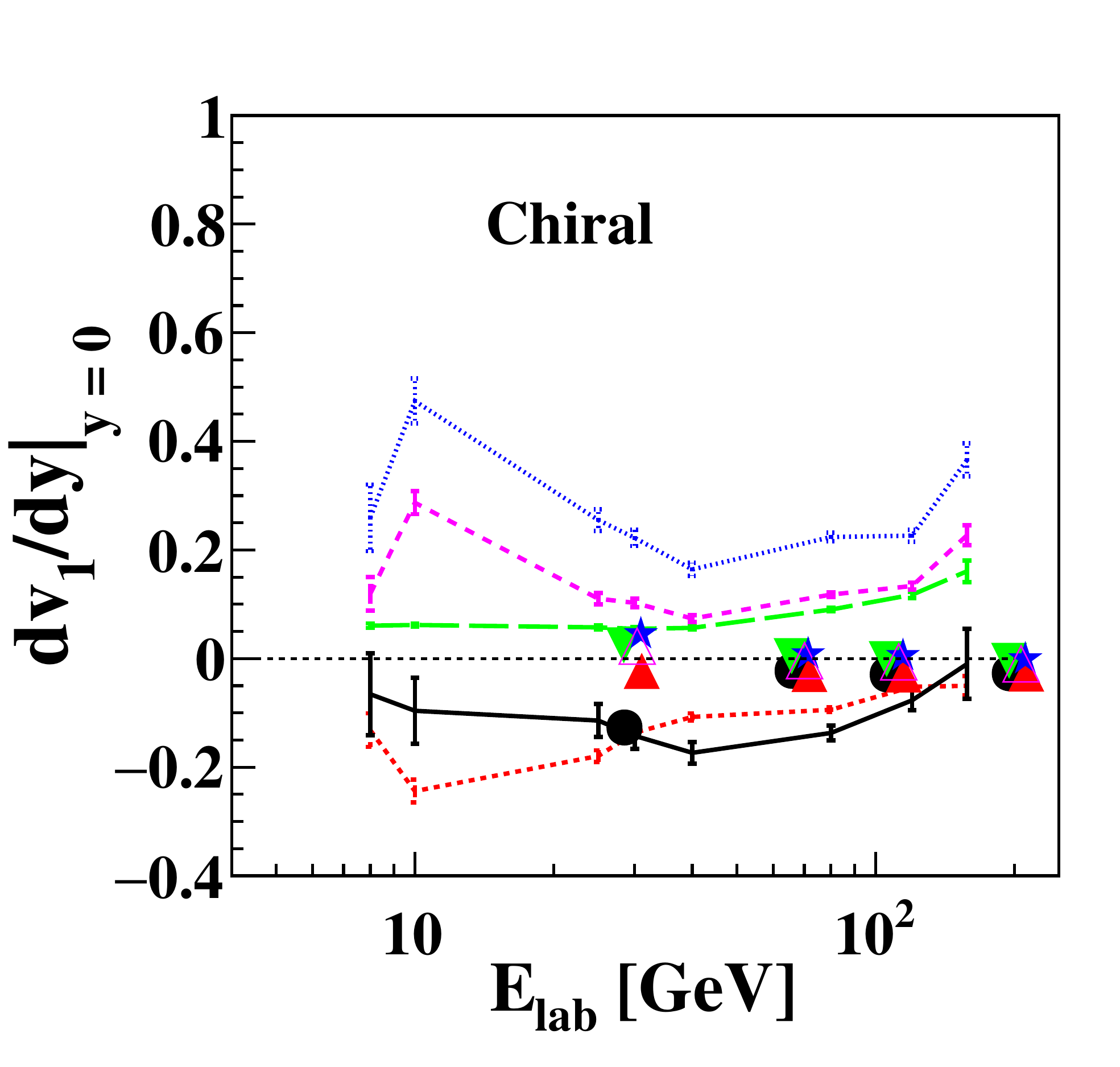}
\includegraphics[scale=0.30]{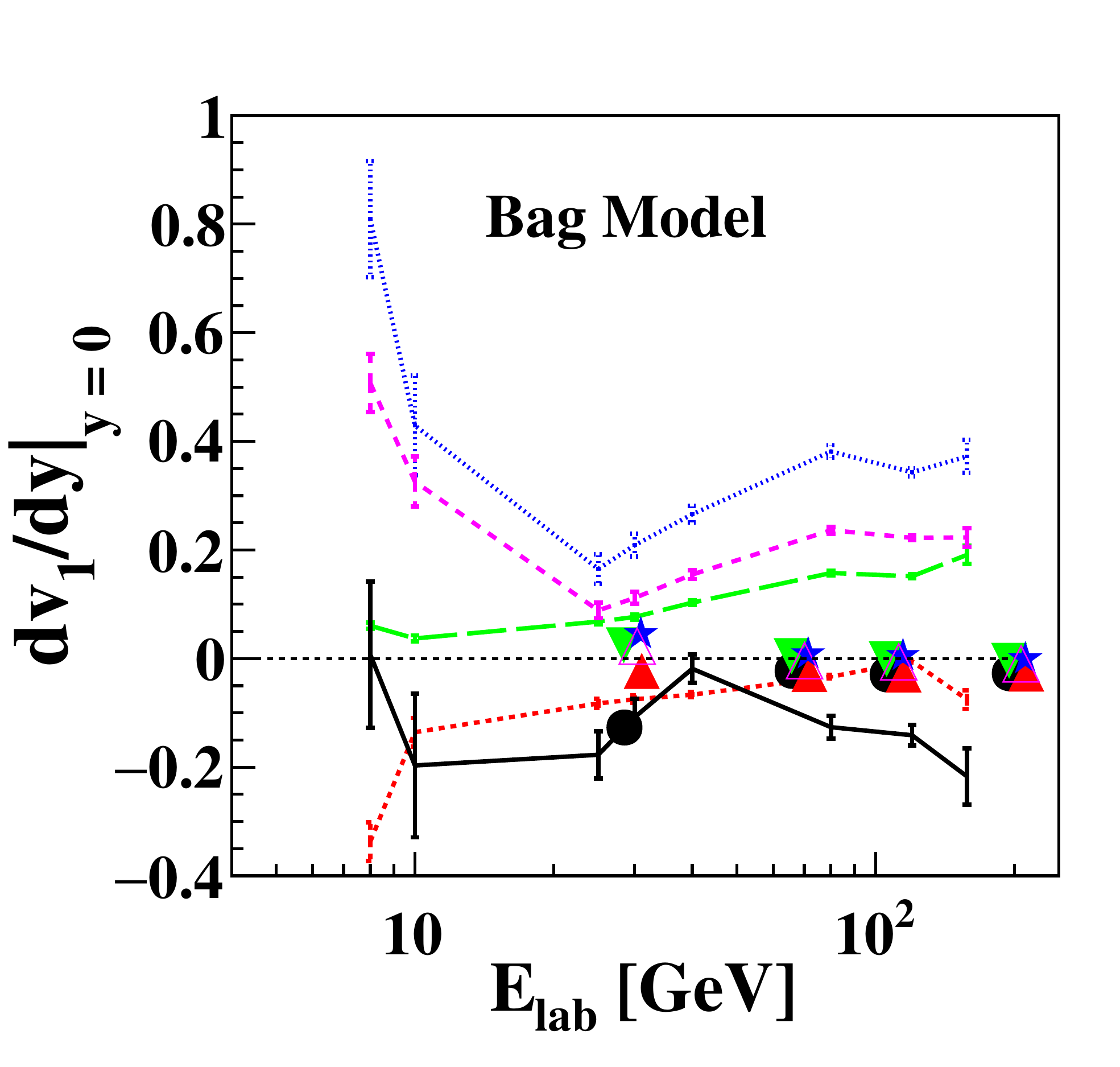}
\includegraphics[scale=0.28]{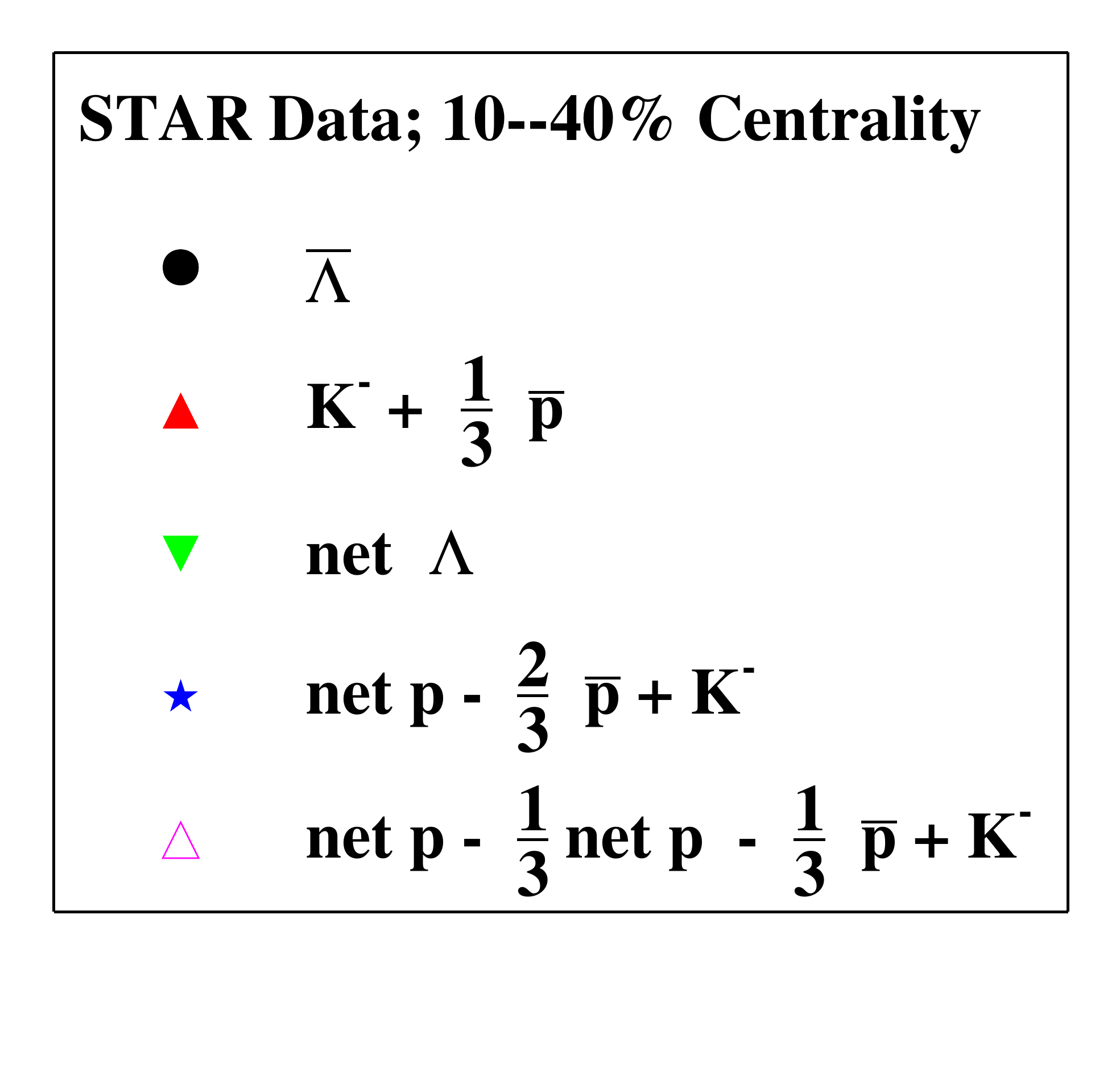}
\end{center}
\caption{Comparison of slope of directed flow of net lambda and anti-lambda with various combinations of hadrons under the assumption of coalescence sum rule as a function of beam energies for different configurations of UrQMD for non-central (b = 5-9 fm corresponds to approximately 10-40$\%$ central) Au-Au collisions. Results from all variants are compared with STAR experimental measurements~\cite{Adamczyk:2017nxg} in 10-40$\%$ central Au-Au collisions. Similar kinematic coverage as in data~\cite{Adamczyk:2017nxg} are applied to the simulations.}
\label{figLam}
\end{figure*}

\section*{III Results and Discussion}

\begin{figure*}
\begin{center}
\includegraphics[scale=0.295]{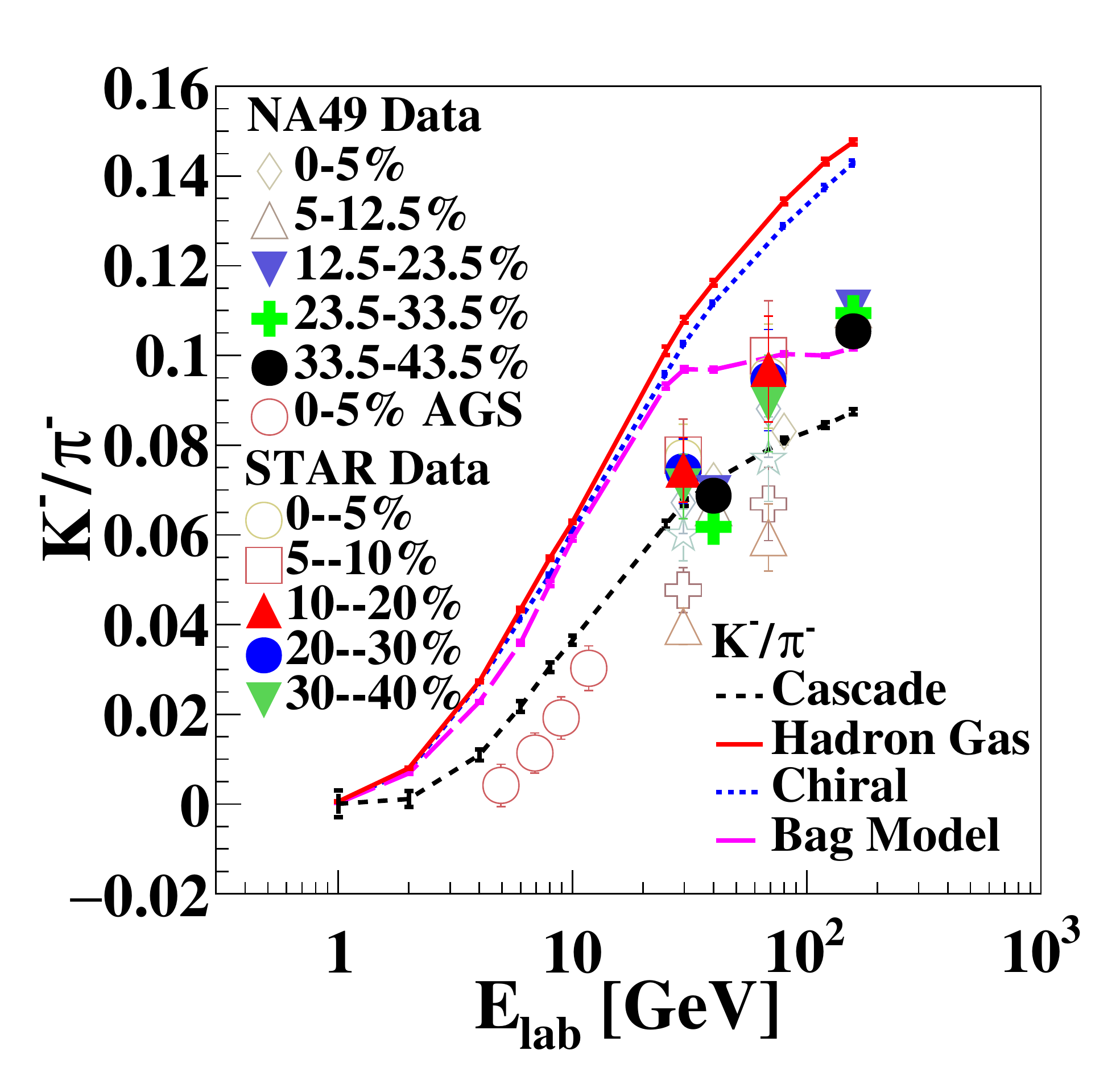}
\includegraphics[scale=0.295]{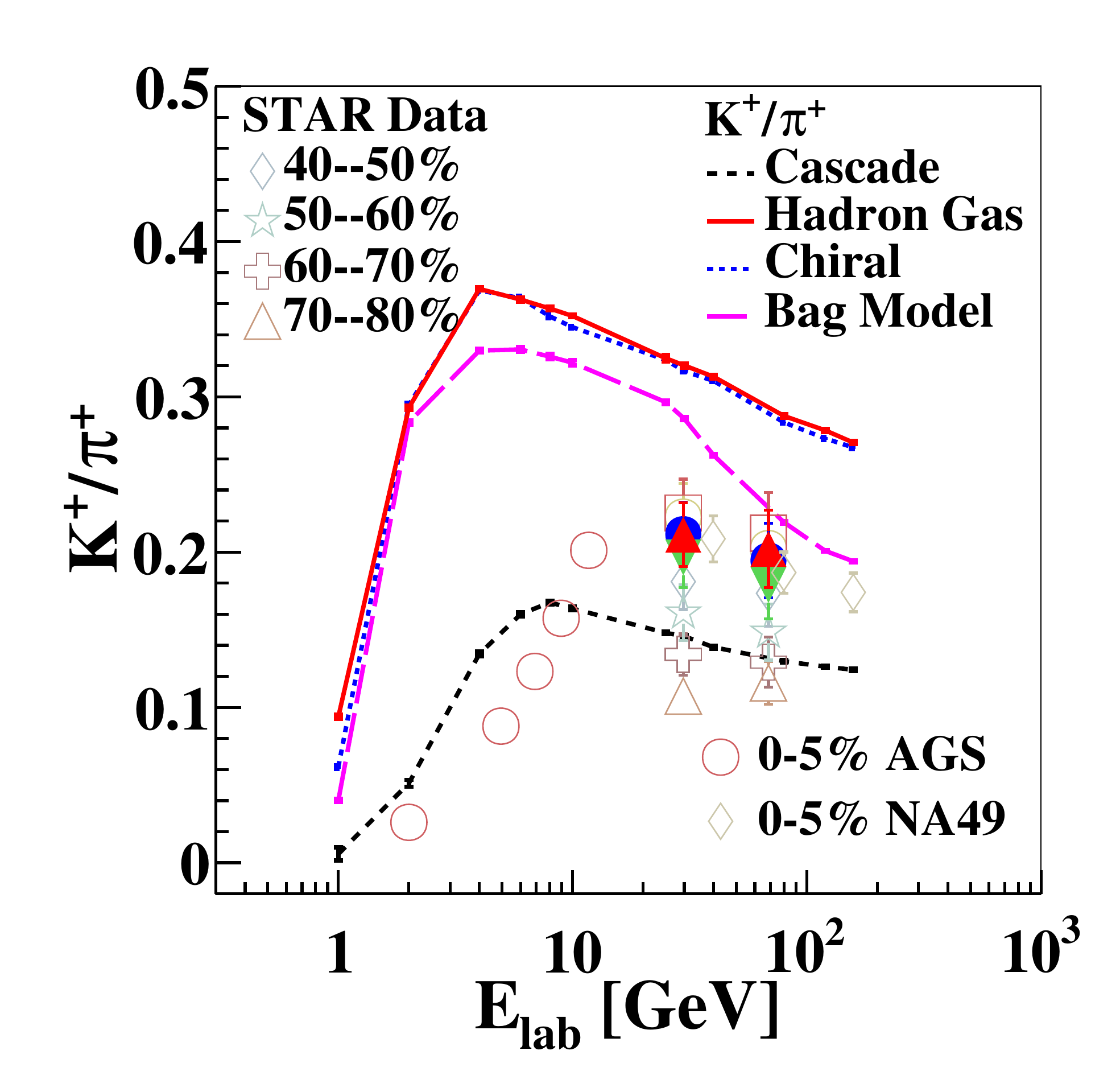}
\includegraphics[scale=0.295]{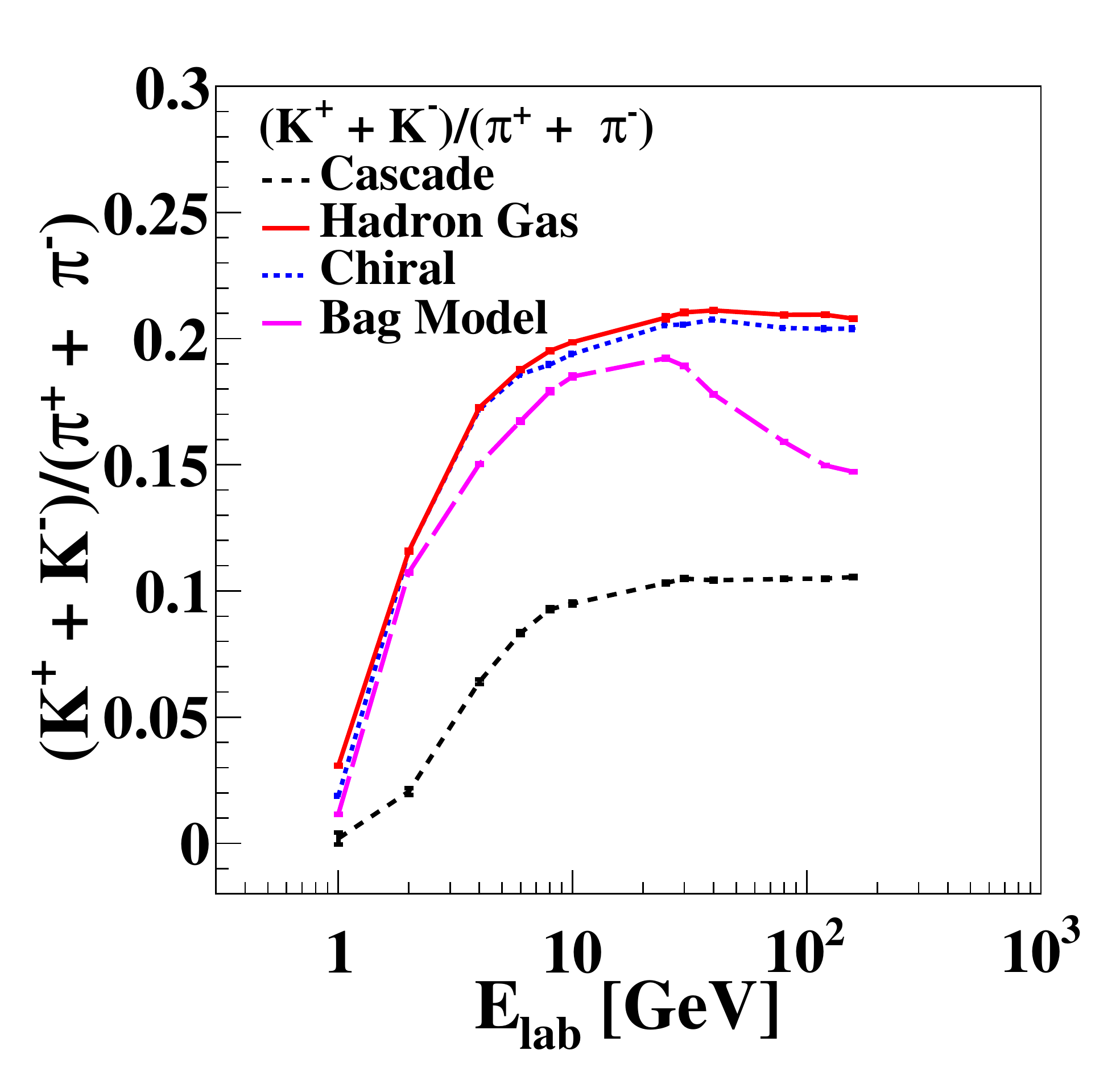}
\end{center}
\caption{$\rm K^{-}$ to $\pi^{-}$, $\rm K^{+}$ to $\pi^{+}$ and ($\rm K^{+}$ + $\rm K^{-}$)/($\pi^{+}$ + $\pi^{-}$) ratio as a function of beam energy for different configurations of UrQMD for non-central (b = 5-9 fm corresponds to approximately 10-40$\%$ central) Au-Au collisions and their comparison with AGS~\cite{E866:1999ktz}, NA49~\cite{NA49:2002pzu,Alt:2006dk} and STAR experimental measurements~\cite{Adamczyk:2017iwn} in Au-Au, Pb-Pb and Au-Au collisions for all available centralities, respectively. Vertical bars on the data denote statistical uncertainties.}
\label{ktopi}
\end{figure*}

\begin{figure*}
\begin{center}
\includegraphics[scale=0.295]{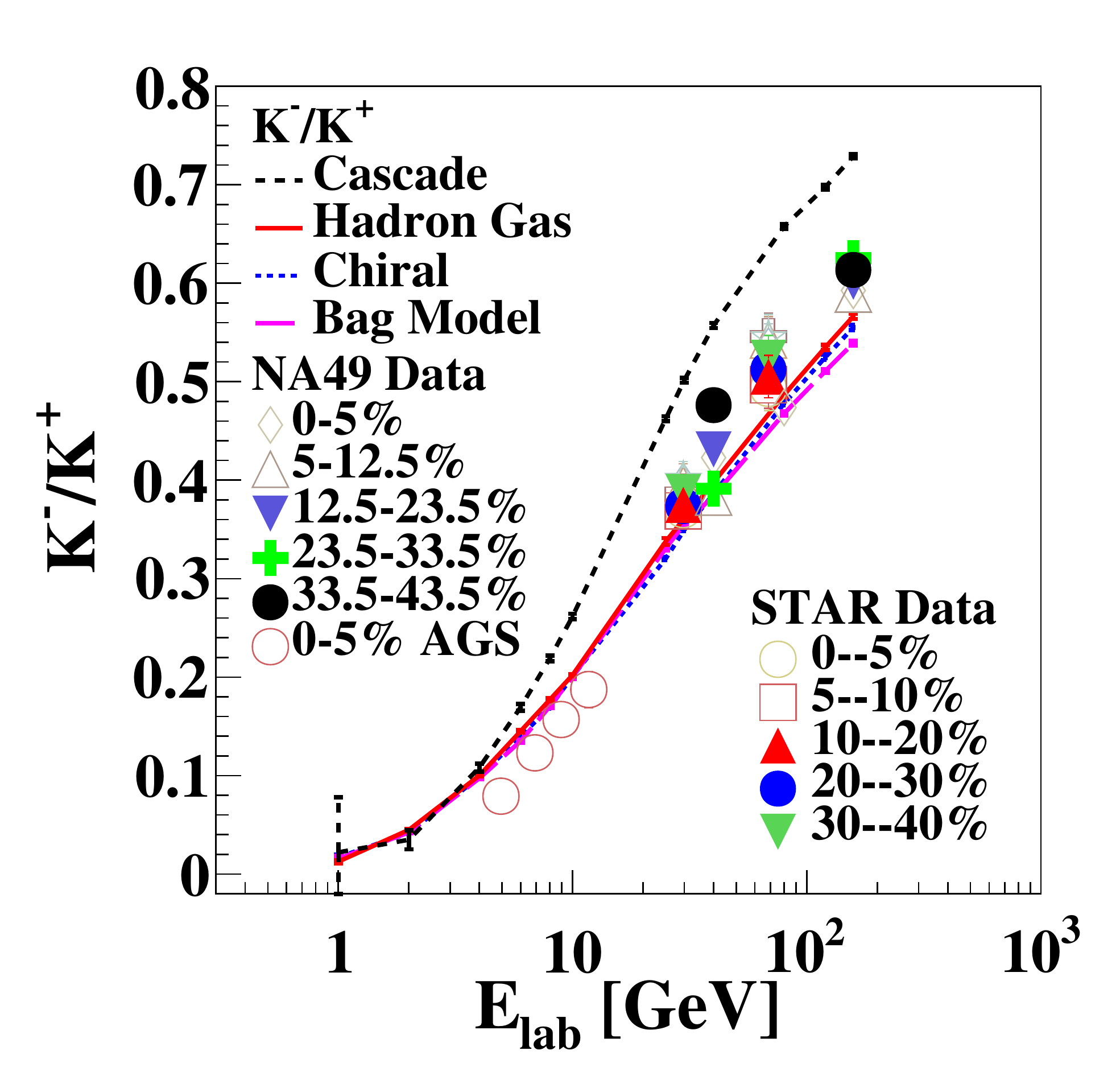}
\includegraphics[scale=0.295]{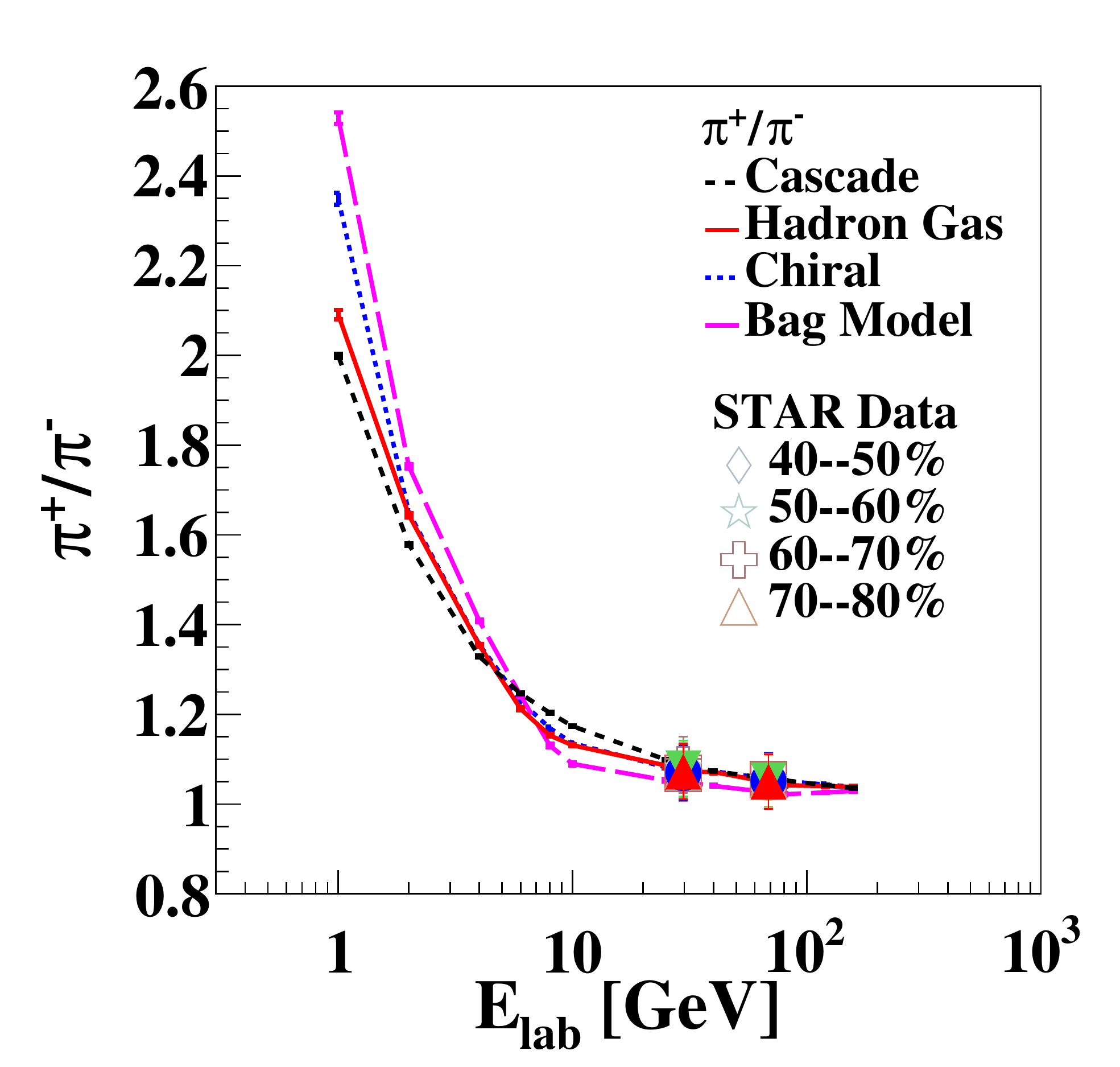}
\includegraphics[scale=0.295]{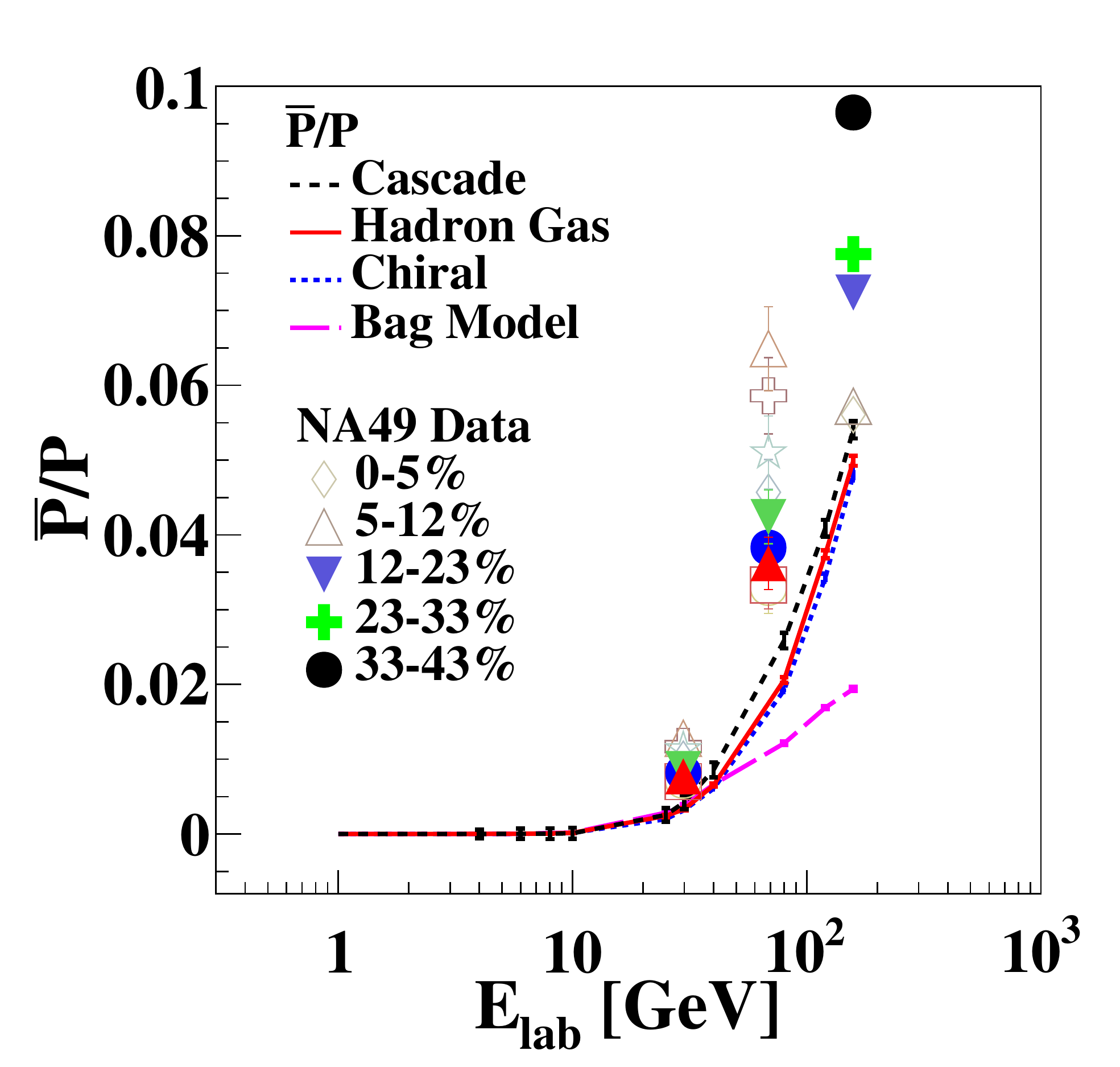}
\end{center}
\caption{$\rm K^{-}$ to $\rm K^{+}$, $\pi^{+}$ to $\pi^{-}$ and anti-proton to proton ratio as a function of beam energy for different configurations of UrQMD for non-central (b = 5-9 fm corresponds to approximately 10-40$\%$ central) Au-Au collisions and their comparison with AGS~\cite{E866:2000dog}, NA49~\cite{Alt:2006dk} and STAR experimental measurements~\cite{Adamczyk:2017iwn} in Au-Au, Pb-Pb and Au-Au collisions for all available centralities, respectively. Vertical bars on the data denote statistical uncertainties.}
\label{antiptop}
\end{figure*}

\begin{figure*}
\begin{center}
\includegraphics[scale=0.3]{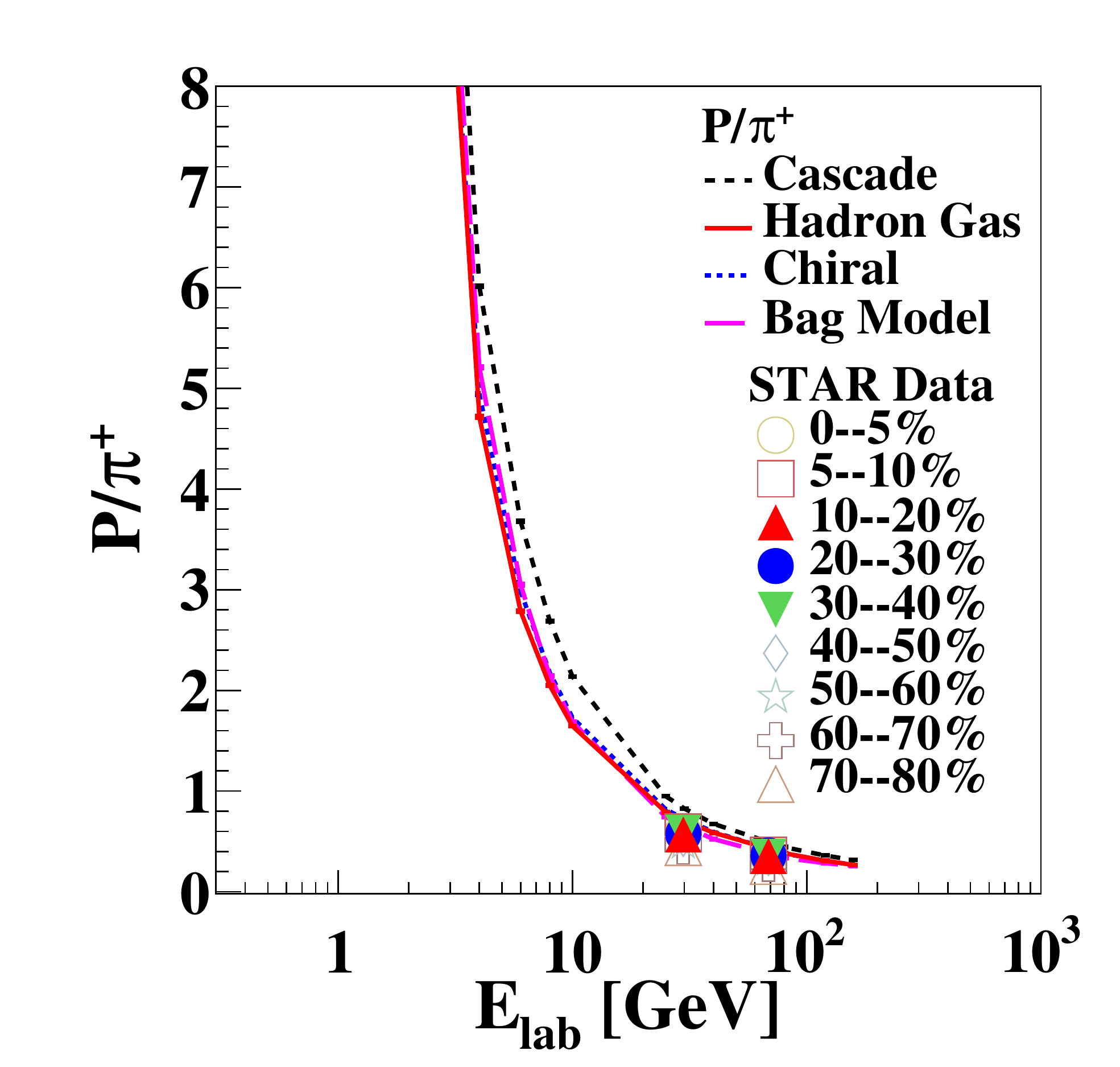}
\includegraphics[scale=0.3]{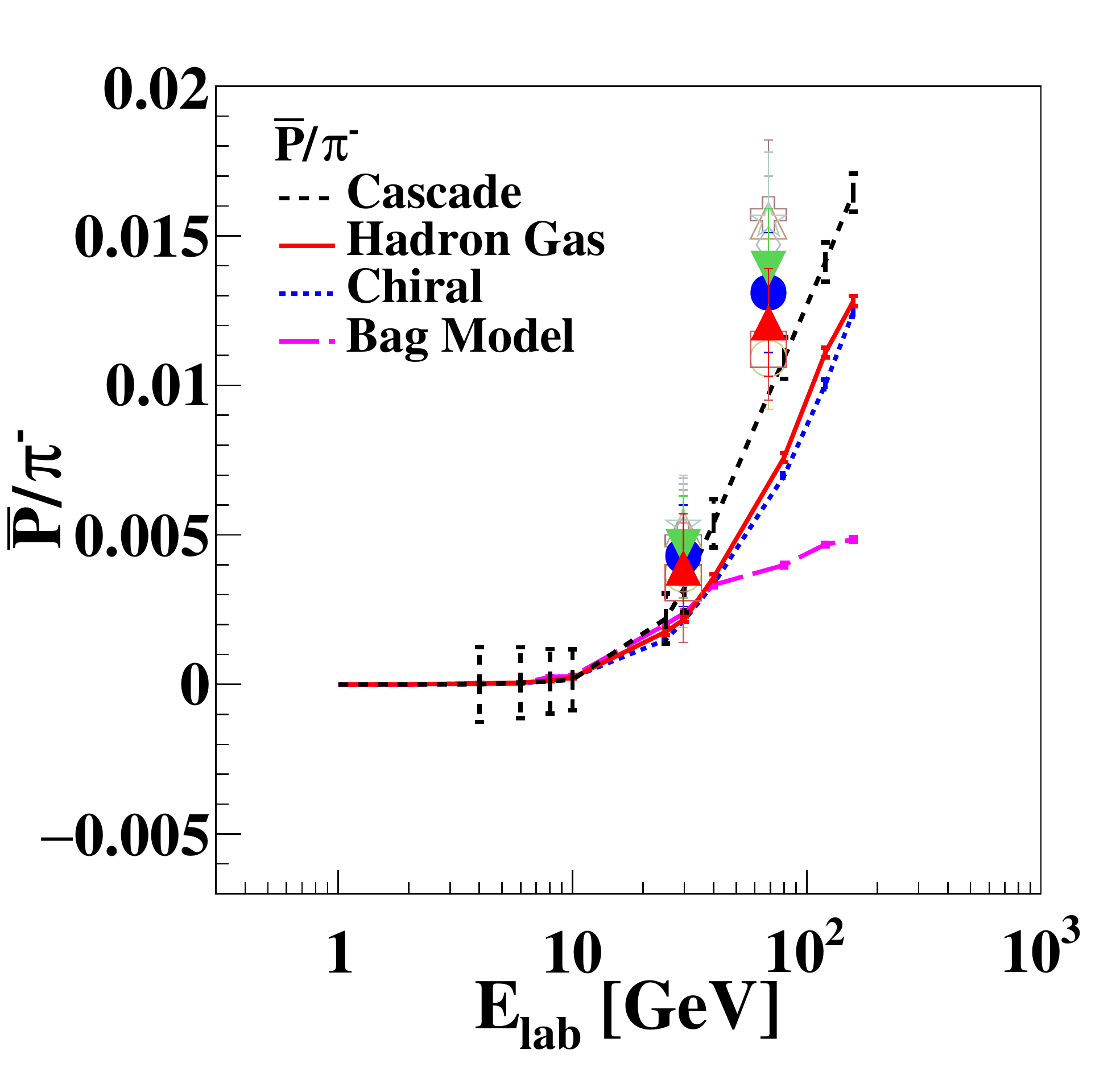}
\end{center}
\caption{Proton to $\pi^{+}$ and anti-proton to $\pi^{-}$ ratio as a function of beam energy for different configurations of UrQMD for non-central (b = 5-9 fm corresponds to approximately 10-40$\%$ central) Au-Au collisions and their comparison with STAR experimental measurements~\cite{Adamczyk:2017iwn} in Au-Au collisions for all available centralities. Vertical bars on the data denote statistical uncertainties.}
\label{ppi}
\end{figure*}

In this section, we present the results of our investigations on various anisotropic flow coefficients at different beam energies for charged and identified hadrons. All the three EoS mentioned above are employed for this purpose. Then we move on to investigate the sensitivity of underlying EoS to the different particle production mechanisms such as strange to non-strange ratio, baryon to meson ratio and so on. Finally, we also look at the net proton rapidity spectra for different EoS to look for possible insights into the longitudinal dynamics of the medium.

\begin{figure*}
\begin{center}
\includegraphics[scale=0.4]{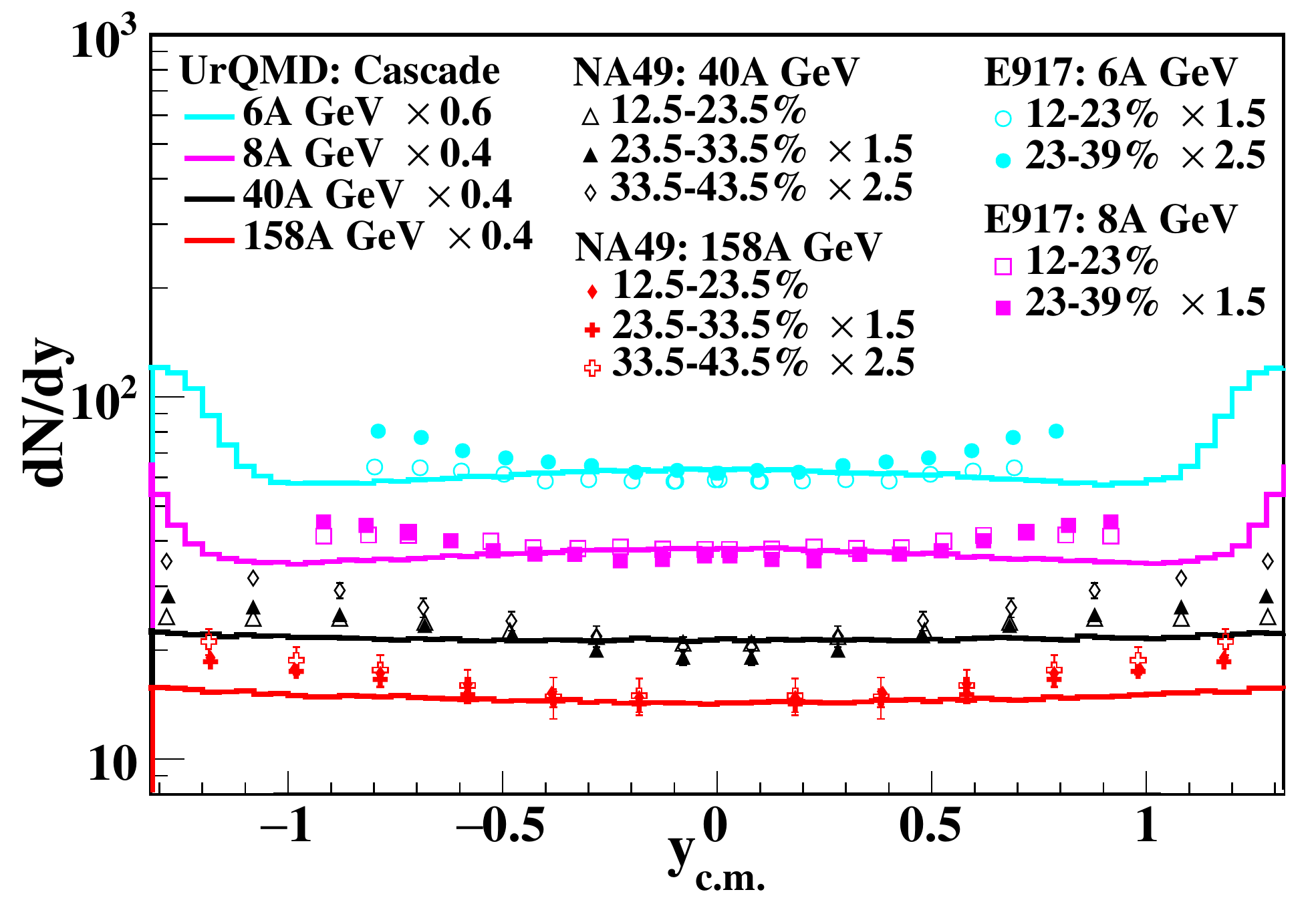}
\includegraphics[scale=0.4]{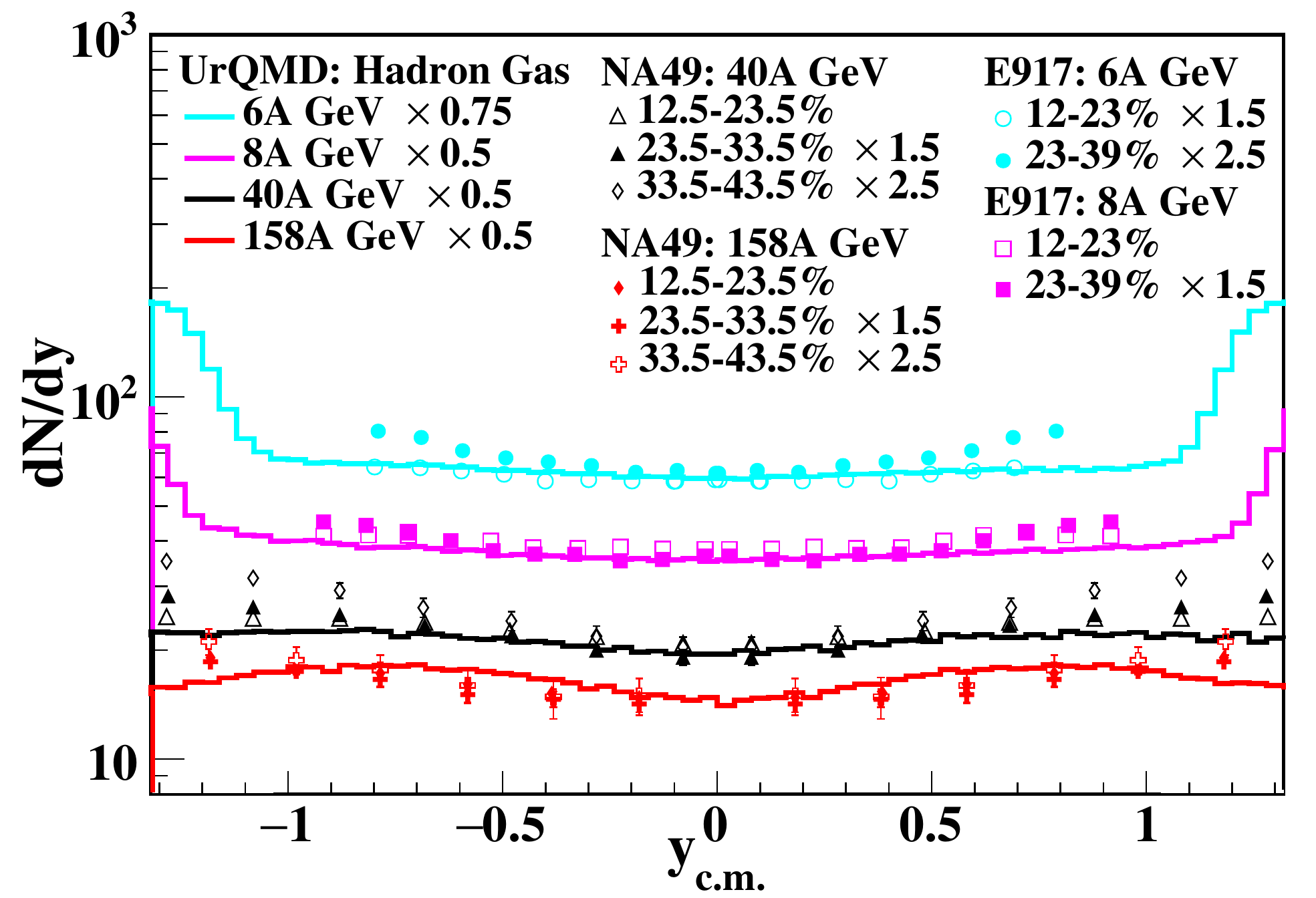}\\
\includegraphics[scale=0.4]{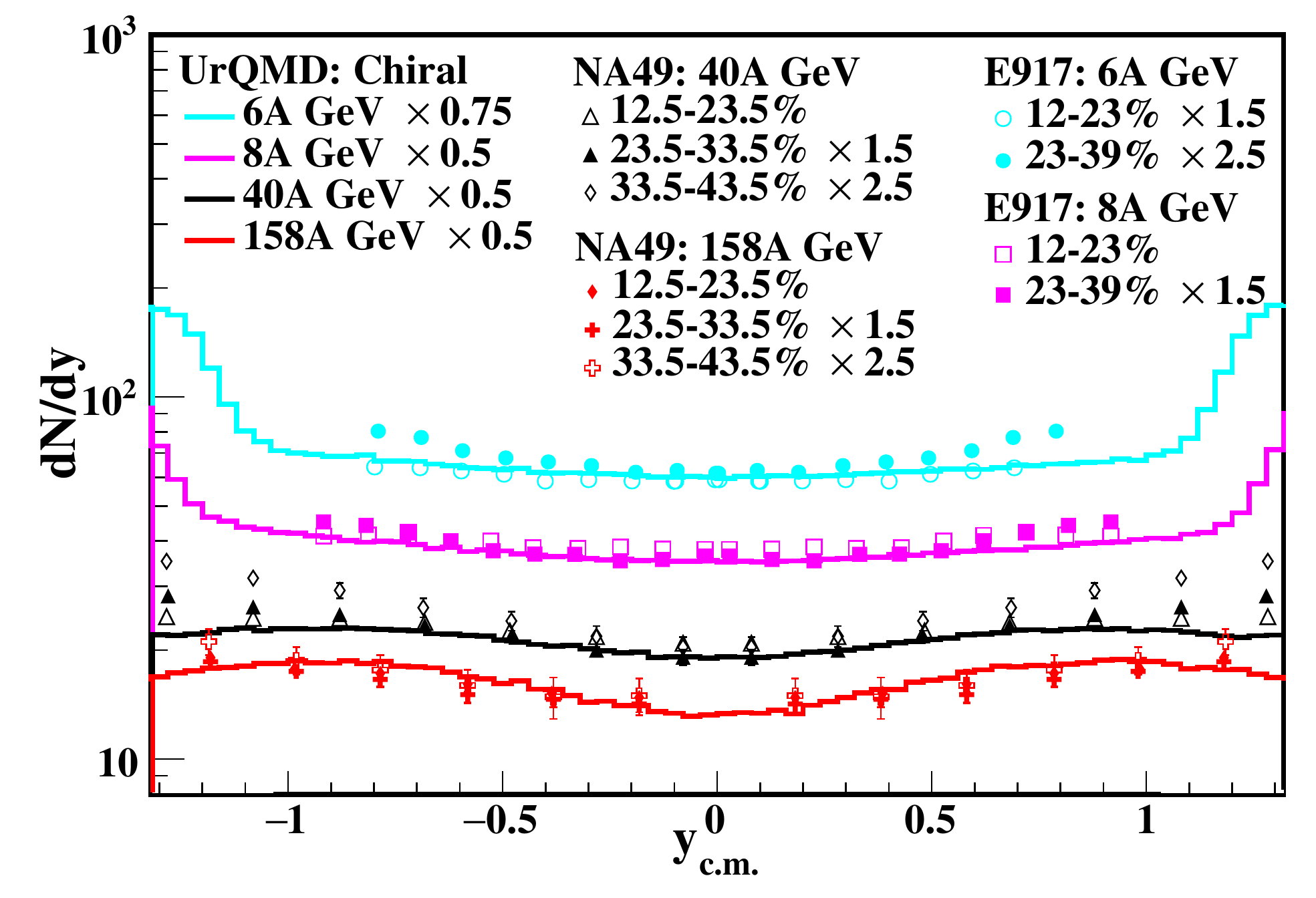}
\includegraphics[scale=0.4]{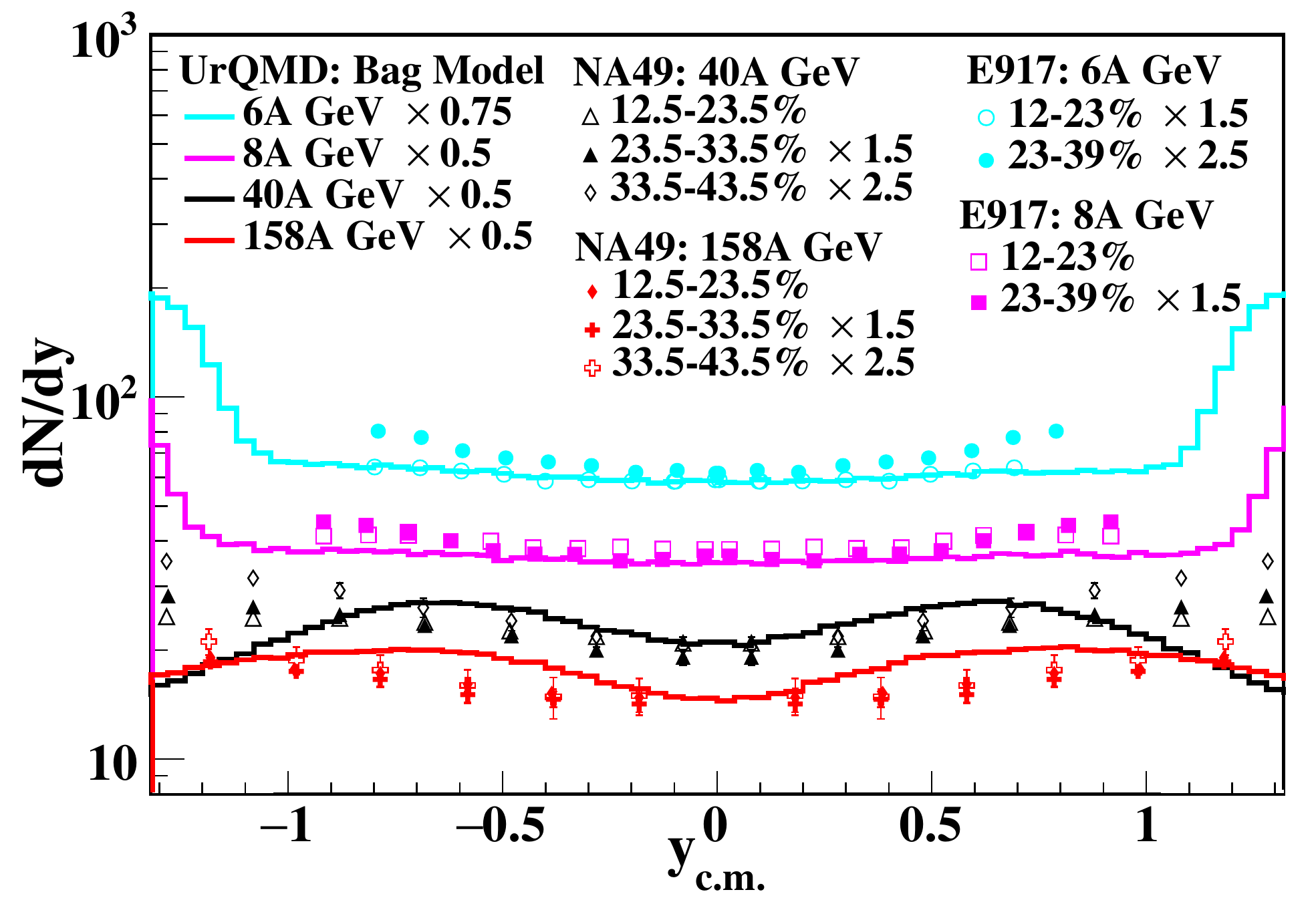}
\end{center}
\caption{Rapidity spectra of net-protons at various beam energies for different equations of state for non-central (b = 5-9 fm corresponds to approximately 10-40$\%$ central) Au-Au collisions and its comparison with the measured rapidity spectra of net-protons in Au-Au and Pb-Pb collisions by E917~\cite{Back:2000ru} and NA49~\cite{Anticic:2010mp} collaborations, respectively. Both simulation results and measurements are scaled for better visualization. Vertical bars on the data denote statistical uncertainties.}
\label{netprap}
\end{figure*}

\begin{figure}
\begin{center}
\includegraphics[scale=0.45]{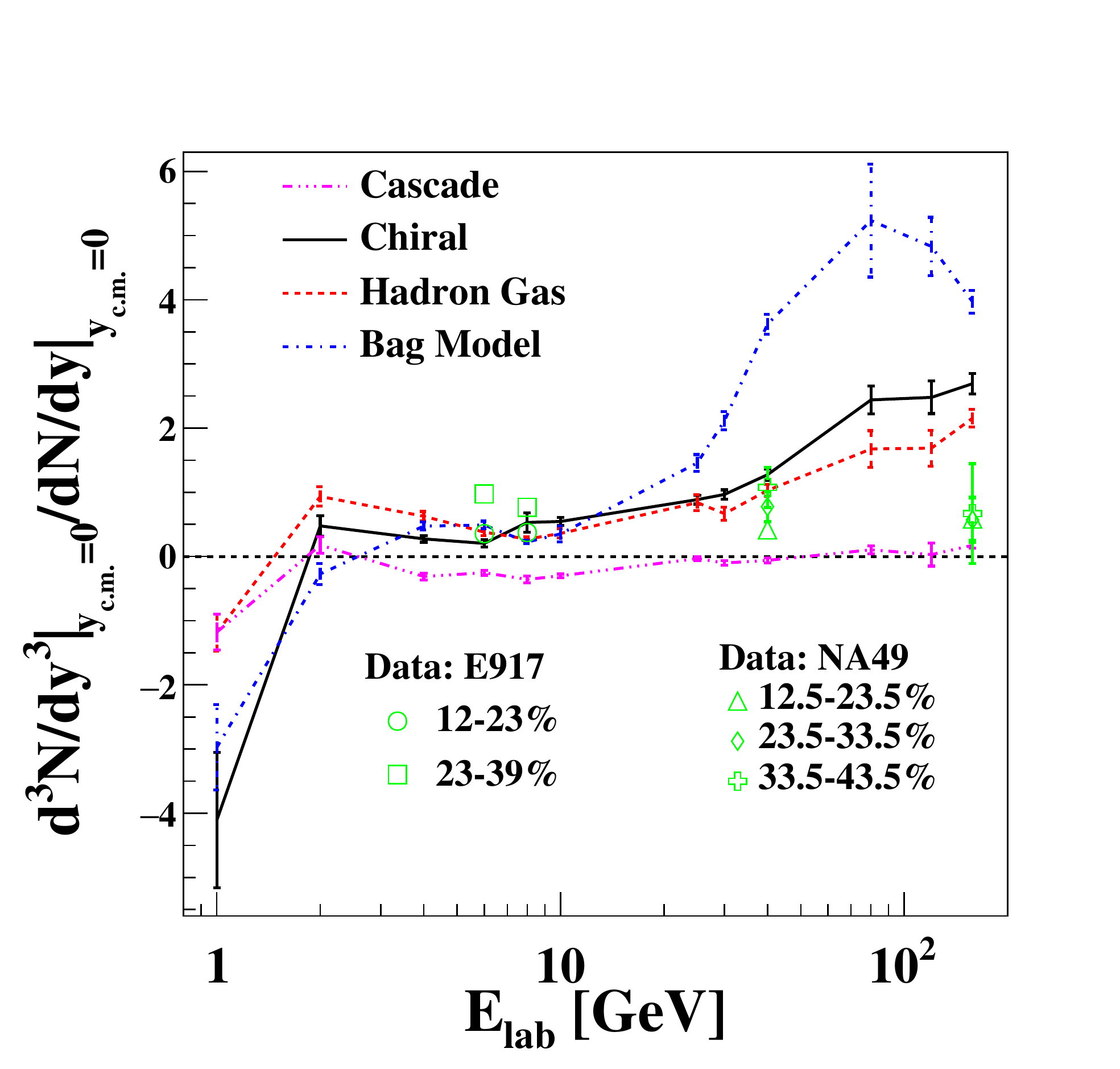}

\end{center}
\caption{Reduced curvature of rapidity spectra of net-protons as a function of beam energy for different configurations of UrQMD for non-central (b = 5-9 fm corresponds to approximately 10-40$\%$ central) Au-Au collisions at midrapidity (-0.5 $<$ $y_{c.m.}$ $<$ 0.5) and its comparison with the calculated reduced curvature of measured rapidity spectra of net-protons in non-central Au-Au and Pb-Pb collisions by E917~\cite{Back:2000ru} and NA49~\cite{Anticic:2010mp} collaborations, respectively.}
\label{bst}
\end{figure}

\subsection*{A. Anisotropic flow coefficients}
Among various harmonic coefficients, $v_{1}$ is believed to hold sensitivity against the longitudinal dynamics of the QCD medium. Therefore, we start by estimating the $v_{1}$ of charged hadrons as a function of rapidity at different beam energies and for pure transport and hybrid versions of the UrQMD model. The results are shown in Fig.~\ref{fig1}. In presence of hydrodynamic expansion, the slope at mid-rapidity remains positive at all investigated energies. For a pure transport approach, the slope initially remains positive and eventually becomes negative. 

Directed flow of pions and protons for $p_{\rm T}<$ 2 GeV/$c$ at 40A and 158A GeV are compared with the existing measurement by NA49 experiment~\cite{Alt:2003ab} at SPS in 10--40$\%$ central Au + Au collisions as shown in Fig.~\ref{fig2}. Hybrid mode fails to explain the slope of $v_{1}$ except for pions at 40A GeV. The pure transport approach is seen to do a better job of explaining the proton $v_{1}$ reasonably well at both energies at midrapidity, an observation in line with previous studies~\cite{Petersen:2006vm,Stocker:2007pd}.

Slopes of directed flow of charged hadrons, pions, protons and net-protons as a function of beam energy are quantified in Fig.~\ref{fig3}. The slope is obtained by fitting differential directed flow ($v_{1}(y)$) using first order polynomial at mid-rapidity. Similar values of slopes are noticed in all three cases of hydro mode up to 10A GeV for all species. The slope using cascade mode is smaller compared to hydro mode. For pions, the slope obtained in cascade model always remain negative at all investigated energies and show transition from negative to positive value between 30A to 80A GeV once hydrodynamic expansion is switched on. The slope does not show any sensitivity to underlying dofs brought by HG and chiral EoS in charged hadrons case which was also observed in our previous study between beam energies 6A--25A GeV~\cite{Rode:2019pey}. Moreover, we tend to see a slight hint of sensitivity in protons and net-protons case beyond 25A GeV, however, we cannot make any strong claim at the moment. In all three EoS cases of hybrid mode, the minimum in slope is observed between 10A--80A GeV. However, in case of Bag Model EoS, the minimum occurs near 10A--25A GeV while for case of other two EoS, the minimum is slightly shifted to higher energy and lies between 25A--80A GeV.  This shift in minimum leads to a splitting of slope parameters of $v_{1}(y)$ between Bag Model and other two EoS which lie around 25--30A GeV. A strong increase of slope in case of Bag Model is observed which could possibly be a result of the in built first order phase transition and perhaps, hint towards the possible onset of deconfinement. In the past, similar interesting feature around similar beam energy has been observed for strange to non-strange ratio~\cite{Alt:2007aa}. Moreover, the slope of directed flow of protons is compared with the available experimental measurements of E895~\cite{Liu:2000am}, NA49~\cite{Alt:2003ab} and STAR~\cite{Adamczyk:2014ipa} collaborations as depicted in Fig.~\ref{fig3}.  It reveals that the results with hybrid mode overestimate the data beyond 2A GeV. Moreover, the slope of $v_{1}$ of pions and net-protons is compared with STAR experiment measurements and it is observed that the inclusion of hydrodynamic expansion overestimates the measurements. According to the fluid dynamical calculations, the slope of $v_{1}$ of the baryons is expected to change sign attributed to softening of EoS in the presence of first order phase transition. This was tested with various freeze-out scenarios using hydrodynamical simulations in Ref.~\cite{Steinheimer:2014pfa}. On the other hand, the results with cascade mode underestimate the measurements below 6A GeV and thereafter, show similar trend with slight overestimation above 30A GeV. 

Moving forward, we attempt to look at the net-protons for $p_{\rm T}<$ 2 GeV/$c$ in more detail by inspecting their $p_{\rm T}$-integrated directed and elliptic flow at midrapidity (-0.5 $<$ $y_{c.m.}$ $<$ 0.5) as a function of beam energies as shown in Fig.~\ref{fig4}. In the left plot, we observe alike trend for $v_{1}$ as its slope in all four cases studied here. Moreover, feature of splitting at 20--30A GeV in presence of hydrodynamical evolution is also observed. While in case of $v_{2}$ in the right plot, we witness a similar splitting between Bag Model EoS and other two EoS. Furthermore, at beam energies around 10--25A GeV a broad peak for $v_{2}$ in case of Bag Model can be seen and for $v_{1}$ as well as its slope, it is appeared as a dip at similar beam energies. We repeat this exercise for $v_{2}$ of kaons and pions for $p_{\rm T}<$ 2 GeV/$c$ as shown in Fig.~\ref{fig5} and here also, kaons and pions confirm the EoS dependent splitting in hydro case, however, the splitting is not prominent in case of pions. Furthermore, the $v_{2}$ of pions and net-protons is also compared with the experimental measurements from E895 and NA49 collaborations~\cite{Pinkenburg:1999ya,Alt:2003ab} and we saw overestimation of the data by hybrid mode here as well.

We now move our focus to look at the higher order flow harmonic coefficient $v_{4}$ which has been argued to be generated under the influence of $4^{th}$ order moment of fluid flow and the intrinsic elliptic flow, $v_{2}$~\cite{Borghini:2005kd,Gombeaud:2009ye,Luzum:2010ae}. Under the assumption of ideal fluid dynamics and without any fluctuations, $v_{2}$ and $v_{4}$ are related to each other as, $v_{4}$ = 0.5$(v_{2})^{2}$. So one can expect to acquire some information about the transport properties of nuclear fireball by estimating the ratio $v_{4}/(v_{2})^{2}$. This ratio has been studied in our previous work~\cite{Rode:2019pey} within beam energy range 6A--25A GeV for different equations of state except Bag model. Prior to this, some phenomenological study has been performed for this observable. In particular, the observations using Parton-Hadron-String Dynamics (PHSD) model~\cite{Konchakovski:2012yg} at different beam energies with Au--Au collisions, have shown the ratio $v_{4}/(v_{2})^{2}$ $\approx$ 2. Moreover, the authors at Ref.~\cite{Nara:2018ijw} have attempted to investigate the enhancement of $v_{4}$ in low energy nuclear collisions using JAM model. Experimentally, the results at RHIC~\cite{Adams:2003zg,Masui:2005aa,Abelev:2007qg,Huang:2008vd} indicated the ratio to be unity. Fig~\ref{figV4} depicts the ratio as a function of beam energy ($\rm E_{\rm Lab}$) for different EoS and the values always remain below 2 for all four cases. The ratio $v_{4}/(v_{2})^{2}$ has been claimed to be in association with the phenomenon of incomplete equilibration in the literature~\cite{Bhalerao:2005mm}. However, the authors have studied this observable as a function of $K^{-1}$, number of collisions per particle. With $K$ being the Knudsen number, a dimensionless quantity and a measure of degree of thermalization, it is a function of system size and beam energy. The local equilibration is expected to be reached when $K^{-1}$ $\gg$ 1. Moreover, the  deviations from ideal hydrodynamics lead to incomplete thermal equilibrium. As the ratio shown in Fig.~\ref{figV4} deviates from 0.5, giving the impression that the system is not fully equilibrated, thus prevent the use of ideal hydrodynamics in these beam energy regimes. The results here can be used to make some robust claims on the degree of thermalization of the nuclear fireball after comparison with the data available from future experiments at FAIR and NICA. 

At last, we attempt to look at the number of constituent quark (NCQ) scaling in the flow coefficients for beam energies examined in this investigation. For this, we specially look at the slope of the directed flow of various species and their combination under the assumption of coalescence sum rule~\cite{Dunlop:2011cf,Adamczyk:2017nxg} for all four variants of UrQMD and the results are shown in Fig~\ref{figLam}. First, similar to Ref.~\cite{Adamczyk:2017nxg},  we compare the $dv_{1}/dy$ values of $\bar{\Lambda}$ ($\overline{uds}$) (black markers) with ($K^{-}$ ($\bar{u}s$) + $\frac{1}{3}\bar{p}$ ($\overline{uud}$)) (red markers) where, the same flow for $s$ and $\bar{s}$ and similarly, for $\bar{u}$ and $\bar{d}$ is assumed. Same kinematic coverage as in experimental measurements are applied in our simulations for all species. Though our results are quantitatively higher than the ones presented in Ref.~\cite{Adamczyk:2017nxg}, however qualitatively, the sum rule seemed to be followed for these two cases at higher beam energies with slight hint of violation below 25A GeV which at the moment, can not be strongly claimed due to large uncertainties in all four cases. For the same reason, we plot our results above 8A GeV upto 158A GeV. Moreover, we also look at one more set which is not as simple as earlier one. As discussed in Ref.~\cite{Adamczyk:2017nxg}, different directed flow for transported\footnotemark \footnotetext[1]{quarks transported from the initial nuclei} and produced\footnotemark \footnotetext[2]{produced in the interactions} quarks is expected which are not easy to distinguish in practice. The comparison of $dv_{1}/dy$ of net $\Lambda$ ($uds$) (blue triangular markers) with the calculation comprising different combinations of net $p$ ($uud$), $\bar{p}$ ($\overline{uud}$) and $K^{-}$ ($\bar{u}s$) (pink circle and blue square markers) is shown in Fig.~\ref{figLam}.  The combination of $K^{-}$ and $\frac{1}{3}$ $\bar{p}$ would give $s$ quark which is assumed to replace produced $u$ quark in net $p$ in the first coalescence calculation (pink circle markers). This calculation is expected to hold true at relatively higher energies where most of the quarks are produced and may not be valid at beam energies considered in this investigation, and it seems to be the case from our observations as evident from Fig.~\ref{figLam} for all four cases of UrQMD. Contrary, in the second calculation where net $p$ is added up with $s$ quark, it is assumed that the transported quarks have dominant contribution in net $p$, which is quite suitable in the limit of low beam energies, and one of the quarks is replaced by $s$ quark. This calculation shows a nice agreement with net $\Lambda$ between 25A--158A GeV which then breaks down below 25A GeV in all four cases. This further may indicate towards possible confinement to deconfinement transition above 25A GeV which has been predicted in prior studies and also in our investigations earlier in this section. The results from our simulations are compared with the available experimental measurements from STAR collaboration~\cite{Adamczyk:2017nxg} and it is noticed that the measurements are overestimated. It is interesting to see the agreement of these sum rule calculations with EoS cases where the underlying degrees of freedom are not partonic and this needs to be understood. However, the scaling behaviour using pure transport UrQMD model has also been found earlier~\cite{Bhaduri:2010wi}.  
This also brings up the question of whether the underlying assumption of coalescence is indeed the source of this agreement. As mentioned earlier, the particle production in UrQMD at higher energies is performed in terms of string excitation and subsequent fragmentations as narrated in Refs.~\cite{Bleicher:1999xi,Guo:2012qi}. As per the string-excitation scheme, the quark-antiquark or diquark-antidiquark pairs are spontaneously formed in color flux tube between initial quarks and subsequently, mesons and baryons are produced. The produced hadrons undergo multiple scatterings, however, no string will be involved after certain energy limit ($\sqrt{s}$ $<$ 5 GeV). This mechanism could give such outcomes shown in Fig.~\ref{figLam}. It is also worth to note that the additive quark model (AQM) is implemented in UrQMD to estimate the unknown hadronic cross-sections~\cite{Bleicher:1999xi}. This model assumes the existence of very weakly interacting dressed valance quarks inside the hadrons.

\subsection*{B. Particle ratios}   
     
In this subsection, we proceed to investigate and understand the effect of different degrees of freedom and phase transition on the yield of the final state particles. For this, we obtain various particle ratios namely, strange to non-strange, anti-particle to particle and compare them with the available data. For central collisions, the $\rm K^{+}/\pi^{+}$ ratio, has been studied in literature as a unique measure of the onset of deconfinement~\cite{Alt:2007aa}. It will be interesting to see the behavior of this observable in case of non-central collisions. We estimate various particle ratios to procure insights about the medium properties by studying the impact of different equations of state. In Fig.~\ref{ktopi}, we show ratio of $\rm K^{-}/\pi^{-}$, $\rm K^{+}/\pi^{+}$ and $\rm (K^{+} + K^{-})/(\pi^{+} + \pi^{-})$ as a function of beam energy for different cases of EoS.  In the left plot, $\rm K^{-}/\pi^{-}$ shows a monotonic rise for all beam energies except for Bag Model EoS which saturates after 20A GeV. In the middle plot, $\rm K^{+}/\pi^{+}$ ratio shows a similar increasing behavior up to 4A GeV and then start to decrease with hint of stronger drop in case of Bag Model EoS between 20A--30A GeV. In the right most plot, $\rm (K^{+} + K^{-})/(\pi^{+} + \pi^{-})$ is obtained as a function of beam energy and similar splitting seen earlier between 20--30A GeV in presence of first order phase transition is observed. The ratio seems to saturate beyond this range in all other scenarios. Due to the unavailability of the measurements in the desired centrality class, both $\rm K^{-}/\pi^{-}$ and $\rm K^{+}/\pi^{+}$ ratios are compared with the experimental data from NA49~\cite{Alt:2006dk} and STAR experiment~\cite{Adamczyk:2017iwn} at three different centralities as these seem to cover impact parameters considered in this work. To demonstrate the centrality and beam energy dependence, we also compare our predictions of various particle ratios with measurements at most central as well as peripheral collisions. From Fig.~\ref{ktopi}, it can be seen that the chiral and hadron gas EoS are able to reproduce the trend set by data in both these ratios however the magnitude is overestimated. Furthermore, we also look at the antiparticle to particle ratio for different EoS. In Fig.~\ref{antiptop}, $\rm K^{+}/K^{-}$, $\rm \pi^{+}/\pi^{-}$ and $\rm \bar{p}/p$ ratios are depicted for all four cases of fireball evolution. $\rm K^{+}/K^{-}$ ratio shows an increase for all beam energies and EoS, however, no sensitivity to a specific EoS is detected.  In the middle plot of Fig~\ref{antiptop}, we see same magnitude of $\rm \pi^{+}/\pi^{-}$ ratio for all EoS at all energies beyond 4A GeV with decreasing trend as a function of beam energy. Moreover, data seem to favor hybrid mode for both ratios with slight underestimation by hybrid in the case of $\rm K^{+}/K^{-}$ ratio. The anti-proton to proton ratio is shown in the right most plot and compared with experimental data.  Measurements are relatively underestimated by the model in all cases of EoS. Finally, we study the $\rm p/\pi^{+}$ and $\rm \bar{p}/\pi^{-}$ ratios and compare them with the available data as shown in Fig.~\ref{ppi}. In the former case, the ratio is inversely proportional to beam energy and shows similar magnitude for all hybrid cases with slightly higher magnitude in cascade case at all beam energies. The ratio using hybrid mode shows good agreement with the experimental measurement as depicted in the left plot.  In the right plot of Fig.~\ref{ppi}, $\rm \bar{p}/\pi^{-}$ ratio shows similar trend as data and sensitivity to first order phase transition. The reader may take note of the fact that for protons the comparison of the UrQMD model calculations with the experimentally measured data are to be accepted with a caveat. In low energy collisions ($E_{lab}$ $\lessapprox$ 10A GeV) the production of light nuclei (d,t,He) has a non-negligible  contribution. The model calculates the so-called primordial nucleons which still contain the contribution of the nucleons bound in the light nuclei. This concerns all observables involving protons. However for anisotropic flow co-efficients the effect of bound proton is considerably reduced because of their independence of proton multiplicity. But it is important for the observables like particle ratios presented in Fig.~\ref{antiptop} and Fig.~\ref{ppi}  involving proton yield, as well net proton rapidity distribution shown in Fig.~\ref{netprap} (net subsection). A consistent way to take the light nuclei into account is by coalescing the final state nucleons from UrQMD. The basic philosophy behind coalescence approach is to check for clusters of nucleons at freeze-out with a very small momentum difference that happens to be very close to each other. However, this is beyond the scope of the present work. 

 \subsection*{C. Net-proton rapidity spectra}

Understanding the in-medium properties of stopped protons by studying their rapidity distributions have been a promising observable. In Refs.~\cite{Ivanov:2010cu,Ivanov:2012bh,Ivanov:2013mxa,Ivanov:2015vna,Ivanov:2016xev}, multiple of studies in this direction has been performed. It has been argued that the irregularities in the distribution of  stopped protons may well be the consequence of onset of deconfinement transition. This occurs due to inherited softest point in the nuclear equations-of-state in the vicinity of a phase transition. Such investigations are generally performed in central collisions however, it is also worth to check this in non-central collisions as well. The shape of rapidity spectra at midrapidity may contain very crucial information about medium and believed to be sensitive to the underlying nuclear equations of state. Therefore, we look at the net-proton rapidity distribution at different beam energies and equations of state.  In Fig.~\ref{netprap}, we show rapidity distribution of net-protons at mid rapidity for all energies and EoS considered in this work. Rapidity spectra remain flat at high beam energies in case of cascade mode in contrast to hybrid mode where it shows a very interesting feature. These results are compared with the measured rapidity spectra at available energies from E917~\cite{Back:2000ru} and NA49~\cite{Anticic:2010mp} experiments in the centrality regions covering investigated range. Our results overestimate the measurements at all available beam energies.

As the irregularities in the shape of rapidity spectra at midrapidity can potentially help in explaining the dynamics of the medium, we quantify the nature of simulated as well as the measured spectra at midrapidity by calculating the double derivative of the rapidity spectra at midrapidity i.e. global minima or maxima as shown in Fig.~\ref{bst}. This quantity is identical to the one obtained in the Refs.~\cite{Ivanov:2010cu,Ivanov:2012bh,Ivanov:2013mxa,Ivanov:2015vna,Ivanov:2016xev} and is referred as reduced curvature. For this, the rapidity distributions of net-protons are fitted with polynomial at the midrapidity for all beam energies and EoS. As shown in Fig.~\ref{bst}, the reduced curvature in case of cascade remains constant and zero for all energies. As soon as the hydrodynamical evolution is introduced, the corresponding observable show some sensitivity as a function of beam energy. Similar to simulations, the reduced curvatures of measured rapidity spectra shown in Fig.~\ref{netprap} are calculated and it is seen that it remains almost flat, however, slightly higher than cascade case at all beam energies. It almost matches with hybrid scenario at 6A, 8A and 40A GeV and is slightly lower in hybrid mode case at 158A GeV. We do not notice the so-called ``peak-dip-peak-dip''  irregularity as seen in the experimental observations and in the central collisions~\cite{Ivanov:2010cu,Ivanov:2012bh,Ivanov:2013mxa,Ivanov:2015vna,Ivanov:2016xev}. It may also be mentioned that in Ref.~\cite{Ivanov:2010cu,Ivanov:2012bh,Ivanov:2013mxa,Ivanov:2015vna,Ivanov:2016xev}, the contribution of nucleons bound in the light nuclei was subtracted from primordial nucleons by means of the coalescence, whereas our results still suffer form the uncertainty due to inclusion of contribution from bound protons, as detailed earlier.  Still, it is interesting to note that this observable has led to show the sensitivity between Chiral and Hadron gas EoS beyond 25A GeV which is the same energy at which we have seen some interesting feature for other observables investigated in this work. The mangitude and slope of reduced curvature is highest for Bag Model and decreases for Chiral to Hadron gas beyond 25A GeV. In the end, it is worth mentioning that the net-protons are the only species for which any sensitivity to the underlying degrees of freedom has been noticed and that to especially for the observables related to longitudinal dynamics such as directed flow and rapidity.

\section*{IV Summary}
In this article, we have made dedicated efforts to understand the impact of various nuclear equations of state on the several observables of the nuclear matter produced in the low energy collisions of heavy-ions in wide range of beam energies, 1A--158A GeV.  The UrQMD model with intermediate hydrodynamical evolution was employed with different nuclear equations-of-state such as Hadron gas, Chiral + deconfinement and Bag model. We started with examining the anisotropic flow coefficients of charged and identified hadrons in above-mentioned beam energy range. A unique feature at 25--30A GeV in the energy dependence of slope of directed flow of charged hadrons, protons and net-protons at mid-rapidity was observed. The slope using Bag Model EoS, showed a splitting, leading to a sharp rise compared to the other two equations-of-state. This may be attributed to the incorporated first order phase transition in the former case. Similar feature was observed for directed flow as well. Apart from the splitting, the dip within certain energy range for these equations-of-state hint towards possible onset of decofinement. Moreover, we noticed that study of net-proton rapidity distribution certainly brings out the sensitivity to underlying degrees of freedom in chiral and hadron gas EoS beyond 20--30A GeV beam energy, however, more evidence in this direction is required to make any robust claim. Along with this, efforts have been made to study the effect of different EoS on elliptic flow ($v_{2}$) of identified hadrons as a function of the beam energy ($\rm E_{\rm Lab}$). As quadrangular flow, $v_{4}$ is believed to be originated from $v_{2}$ and $4^{th}$ order moment of the fluid flow, the ratio $v_{4}/(v_{2})^{2}$ was examined for wide range of beam energies and different EoS. The ratio is always found to be below 2 for all four cases of EoS and can be tested against the data from the future experiments at NICA~\cite{Kekelidze:2016wkp} and FAIR~\cite{Ablyazimov:2017guv,Sturm:2010yit}.

In addition, we have studied NCQ scaling in terms of coalescence sum rule for slope of directed flow of $\bar{\Lambda}$ and net $\Lambda$. For this purpose, we used different calculations used in Ref.~\cite{Adamczyk:2017nxg} and compared them with $dv_{1}/dy$ of $\bar{\Lambda}$ and net $\Lambda$. Results qualitatively match the expectations for all four variants of UrQMD. This study may also hint towards possible onset of deconfinement at certain beam energy above 25A GeV. Furthermore, it was interesting to notice similar nature of the results even for pure transport and hadrons gas EoS cases where quarks and gluons are not underlying degrees of freedom. 

At this point, it appears from the results of collective flow excitation functions that neither of the EoS is suitable enough to quantitatively reproduce the experimental measurements. Moreover, the nature of the matter might be partonic, however, it does not evolve as a non-viscous ideal fluid as implemented in the present version of the model. Furthermore, higher values of calculated flow coefficients corresponding to data suggest larger pressure gradient in ideal hydrodynamic scenario and therefore, one possibility would be to use viscous hydrodynamics instead of ideal one to account for dissipative effects. Other reasons for the disagreement might be inapplicability of hydrodynamics at low beam energies where transport approach seems to give better agreement and also the fact that reaction plane angle which leads to event by event fluctuations are not taken into account in UrQMD.

Various particle ratios are calculated for all EoS and studied as a function of beam energy. The ratios found to be sensitive to first order phase transition and exhibited different behavior in comparison to other cases. UrQMD including fluid dynamic simulations with Hadron Gas and Chiral EoS is qualitatively able to explain the measured strange to non-strange ratio and overestimate the measurements. However, calculated strange to non-strange ratio showed some interesting features in response to various EoS beyond 25A GeV. Similarly, particle to anti-particle ratios are qualitatively described by hybrid mode with underestimation with respect to the data except for $\pi^{+}/\pi^{-}$ ratio which is nicely explained. Moreover, measured ratio of proton to $\pi^{+}$ is also well described by the predictions, however, calculated anti-proton to $\pi^{-}$ ratio underestimates the data.

We wrap up by studying the rapidity spectra of net-protons for different EoS at various beam energies. The shape of these spectra at mid-rapidity, quantified as a reduced curvature is seen to be sensitive to underlying EoS and shows larger value in case of Bag Model EoS beyond 25A GeV. It also revealed the sensitivity to the underlying degrees of freedom beyond 25A GeV. These investigations provide an opportunity to understand the behavior of the various observables under different nuclear equations-of-state and to compare the results with the outcomes from future experiments. 

\section*{Acknowledgements}
S. K. K. acknowledges the financial support provided by CSIR, New Delhi. We acknowledge the computing facility provided by the Grid Computing Facility at VECC-Kolkata, India.

\end{document}